\documentclass[twocolumn]{aastex701}

\usepackage[dvipsnames]{xcolor}
\usepackage{amsmath}
\newif\ifcomments
\commentstrue

\newcommand{\sersic}{S\'ersic}

\begin{document}

\title{JADES: A Prominent Galaxy Overdensity Candidate within the First 500 Myr}

\author[0000-0002-8876-5248]{Zihao Wu}
\affiliation{Center for Astrophysics $|$ Harvard \& Smithsonian, 60 Garden St., Cambridge MA 02138 USA}
\email[show]{zihao.wu@cfa.harvard.edu}

\author[orcid=0000-0002-2929-3121]{Daniel J.\ Eisenstein}
\affiliation{Center for Astrophysics $|$ Harvard \& Smithsonian, 60 Garden St., Cambridge MA 02138 USA}
\email{deisenstein@cfa.harvard.edu}  

\author[orcid=0000-0002-9280-7594]{Benjamin D.\ Johnson}
\affiliation{Center for Astrophysics $|$ Harvard \& Smithsonian, 60 Garden St., Cambridge MA 02138 USA}
\email{benjamin.johnson@cfa.harvard.edu}  

\author[orcid=0000-0003-4565-8239]{Kevin Hainline}
\affiliation{Steward Observatory, University of Arizona, 933 N. Cherry Avenue, Tucson, AZ 85721, USA}
\email{kevinhainline@arizona.edu} 

\author[orcid=0000-0003-0215-1104]{William M. Baker}
\affiliation{DARK, Niels Bohr Institute, University of Copenhagen, Jagtvej 155A, DK-2200 Copenhagen, Denmark}
\email{william.baker@nbi.ku.dk} 

\author[orcid=0000-0002-8651-9879]{Andrew J.\ Bunker}
\affiliation{Department of Physics, University of Oxford, Denys Wilkinson Building, Keble Road, Oxford OX1 3RH, UK}
\email{andy.bunker@physics.ox.ac.uk} 

\author[0000-0002-0450-7306]{Alex J.\ Cameron}
\affiliation{Cosmic Dawn Center (DAWN), Copenhagen, Denmark}
\affiliation{Niels Bohr Institute, University of Copenhagen, Jagtvej 128, DK-2200, Copenhagen, Denmark}
\email{alex.cameron@nbi.ku.dk} 

\author[orcid=0000-0002-9551-0534]{Emma Curtis-Lake}
\affiliation{Centre for Astrophysics Research, Department of Physics, Astronomy and Mathematics, University of Hertfordshire, Hatfield AL10 9AB, UK}
\email{e.curtis-lake@herts.ac.uk} 

\author[orcid=0000-0002-9708-9958]{A. Lola Danhaive}
\affiliation{Kavli Institute for Cosmology, University of Cambridge, Madingley Road, Cambridge, CB3 0HA, UK}
\affiliation{Cavendish Laboratory, University of Cambridge, 19 JJ Thomson Avenue, Cambridge, CB3 0HE, UK}
\email{ald66@cam.ac.uk} 

\author[orcid=0000-0002-8543-761X]{Ryan Hausen}
\affiliation{Department of Physics and Astronomy, The Johns Hopkins University, 3400 N. Charles St., Baltimore, MD 21218}
\email{rhausen@ucsc.edu} 

\author[orcid=0000-0003-4337-6211]{Jakob M. Helton}
\affiliation{Department of Astronomy \& Astrophysics, The Pennsylvania State University, University Park, PA 16802, USA}
\email{jakobhelton@psu.edu} 

\author[orcid=0000-0001-7673-2257]{Zhiyuan Ji}
\affiliation{Steward Observatory, University of Arizona, 933 N. Cherry Avenue, Tucson, AZ 85721, USA}
\email{zhiyuanji@arizona.edu} 

\author[0000-0002-3642-2446]{Tobias J.\ Looser}
\affiliation{Center for Astrophysics $|$ Harvard \& Smithsonian, 60 Garden St., Cambridge MA 02138 USA}
\email{tobias.looser@cfa.harvard.edu} 

\author[orcid=0000-0002-4985-3819]{Roberto Maiolino}
\affiliation{Kavli Institute for Cosmology, University of Cambridge, Madingley Road, Cambridge, CB3 0HA, UK}
\affiliation{Cavendish Laboratory, University of Cambridge, 19 JJ Thomson Avenue, Cambridge, CB3 0HE, UK}
\affiliation{Department of Physics and Astronomy, University College London, Gower Street, London WC1E 6BT, UK}
\email{rm665@cam.ac.uk} 

\author[0009-0006-4365-2246]{Petra Mengistu}
\affiliation{Department of Astronomy and Astrophysics, University of California, Santa Cruz, 1156 High Street, Santa Cruz, CA 95064, USA}
\email{pmengist@ucsc.edu}

\author[orcid=0000-0002-5104-8245]{Pierluigi Rinaldi}
\affiliation{Space Telescope Science Institute, 3700 San Martin Drive, Baltimore, Maryland 21218, USA}
\email{prinaldi@stsci.edu} 

\author[orcid=0000-0002-4271-0364]{Brant E. Robertson}
\affiliation{Department of Astronomy and Astrophysics, University of California, Santa Cruz, 1156 High Street, Santa Cruz, CA 95064, USA}
\email{brant@ucsc.edu}  

\author[orcid=0000-0002-4622-6617]{Fengwu Sun}
\affiliation{Center for Astrophysics $|$ Harvard \& Smithsonian, 60 Garden St., Cambridge MA 02138 USA}
\email{fengwu.sun@cfa.harvard.edu} 

\author[orcid=0000-0002-8224-4505]{Sandro Tacchella}
\affiliation{Kavli Institute for Cosmology, University of Cambridge, Madingley Road, Cambridge, CB3 0HA, UK}
\affiliation{Cavendish Laboratory, University of Cambridge, 19 JJ Thomson Avenue, Cambridge, CB3 0HE, UK}
\email{st578@cam.ac.uk}  

\author[orcid=0000-0002-9081-2111]{James A. A. Trussler}
\affiliation{Center for Astrophysics $|$ Harvard \& Smithsonian, 60 Garden St., Cambridge MA 02138 USA}
\email{james.trussler@cfa.harvard.edu}

\author[orcid=0000-0003-2919-7495]{Christina C. Williams}
\affiliation{NSF–DOE Vera C. Rubin Observatory/NSF NOIRLab, 950 N. Cherry Ave., Tucson, AZ 85719, USA}
\email{christina.williams@noirlab.edu} 

\author[orcid=0000-0001-9262-9997]{Christopher N. A. Willmer}
\affiliation{Steward Observatory, University of Arizona, 933 N. Cherry Avenue, Tucson, AZ 85721, USA}
\email{cnaw@as.arizona.edu} 

\author[0000-0002-7595-121X]{Joris Witstok}
\affiliation{Cosmic Dawn Center (DAWN), Copenhagen, Denmark}
\affiliation{Niels Bohr Institute, University of Copenhagen, Jagtvej 128, DK-2200, Copenhagen, Denmark}
\email{joris.witstok@nbi.ku.dk}

\begin{abstract}
We report a galaxy overdensity candidate at $z\approx10.5$ in the JWST Advanced Deep Extragalactic Survey (JADES). This overdensity contains 18 galaxies with consistent photometric redshifts within $8$ comoving Mpc in projection. The galaxy number density is four times higher than the field expectation, accounting for one-third of comparably bright galaxies and nearly 50\% of the total star formation rate at $10<z_\mathrm{phot}<12$  in the GOODS--S field. Galaxies in the overdensity more frequently have close companions or substructure, with one-third showing such features within 1\,kpc at consistent photometric redshifts, implying enhanced interactions.  Most galaxies have stellar masses of 0.6--3$\times10^8\,M_\odot$, half-light radii of $\sim200$\,pc, and star formation rates (SFRs) of $\sim5$\,$M_\odot\,\mathrm{yr^{-1}}$. Their stellar masses and SFRs are slightly higher than those of field galaxies, but remain broadly consistent with  typical high-redshift scaling relations. Two compact objects show possible Balmer breaks, suggestive of evolved stellar populations or little red dots (LRDs).  We find tentative evidence for a spatially varying Ly$\alpha$ transmission inferred photometrically, consistent with an emerging ionized bubble. This overdensity provides a rare opportunity for probing the environmental impact on galaxy evolution and the onset of cosmic reionization within the first 500\,Myr.
\end{abstract}

\section{Introduction} 
\label{sec:intro}

Galaxies form and evolve within the cosmic large-scale structure, which consists of filaments, nodes, voids, and walls \citep{Kravtsov2012ARA&A, overzier_realm_2016}. The nodes of the cosmic web represent the most extreme matter overdensities and host the formation of galaxy clusters. Despite the remarkable number of galaxies discovered at $z > 10$  \citep[e.g.,][]{curtis-lake_spectroscopic_2023, carniani_2024, castellano_jwst_2024, Witstok2025GSz13Lya, Naidu2026}, none have yet been observed within a significant overdensity at such early epochs. 

Galaxy overdensities are regarded as progenitors of galaxy clusters. They are often referred to as protoclusters when considered dense enough to eventually collapse into a galaxy cluster, although this designation is mainly practical in simulation contexts \citep{overzier_realm_2016}. At high redshifts, galaxy overdensities are not entirely gravitationally bound, as their central halos have not grown massive enough to dominate the gravitational potential \citep{wechsler_connection_2018}. Their structures therefore trace peaks of the primordial density field \citep{Bardeen1986ApJ}.

Overdensities are the primary sites of star formation in the early Universe. Owing to halo bias \citep{Mo1996}, massive dark matter halos preferentially reside in clustered environments \citep{Chiang2017ApJ, wechsler_connection_2018}. Consequently, overdensities tend to host the most massive galaxies and dominate early stellar mass assembly \citep{Thomas2023}. They contribute $\sim50$\% of the cosmic star formation rate at $z \gtrsim 5$ \citep{Chiang2017ApJ, Lim2024MNRAS, Sun2024ApJ}.

Environmental effects further accelerate galaxy formation and evolution in overdensities \citep{Morishita2025overdensity, Witten2025protocluster, Baker2026A&A}. Star formation in overdensities is enhanced at early epochs due to higher gas inflow \citep{Helton2024ApJEvolved, Li2025metal, Morishita2025overdensity}. This enhanced star formation further facilitates metal enrichment \citep{Li2025metal}, although it may be diluted by accretion of metal-poor gas \citep{Li2022ApJmetal, Wang2022metal, Wang2023metal, Morishita2025ApJ}. The overdense environment also fosters galaxy mergers and interactions, accelerating the buildup of stellar mass \citep{behroozi_universemachine_2019}.

Overdensities play an important role in cosmic reionization \citep{Saxena2023Lya,Witstok2024A&A, Witstok2025MNRAS, Chen2026ApJ}. Reionization is thought to proceed through the growth and percolation of ionized bubbles produced by early galaxies \citep{zahn_simulations_2007}. The detection of Ly$\alpha$ emission at $z = 13$ suggests that such ionized bubbles are already in place at this epoch \citep{Witstok2025GSz13Lya}. However, a tension arises regarding the role of overdensities in cosmic reionization: on the one hand, they host more ionizing sources and thus may produce the largest ionizing bubbles; on the other hand, the higher gas density increases hydrogen recombination rates, which may inhibit ionization \citep{Furlanetto2004ApJ, iliev_simulating_2006, Mesinger2007reionization, kulkarni_reionization_2011, Castellano2016ApJ,Endsley2022MNRAS,Larson2022ApJ,Leonova2022MNRAS,Lu2024MNRAS, Neyer2024, Almualla2025thesan}. 

Several overdensities have been spectroscopically identified up to $z\approx8$ \citep[e.g.,][]{Ishigaki2016ApJ,Hu2021NatAs, Laporte2022A&A, Hashimoto2023ApJ, Morishita2023ApJ, Helton2024overdense,Fudamoto2025arXiv2, Fudamoto2025arXiv,  Li2025MNRAS,Morishita2025ApJ,Witten2025arXiv, Li2026MNRAS}, indicating that proto-cluster environments were already in place at these epochs. At $z>10$, however, overdensities remain rare despite the growing number of individual galaxy detections. The only candidate is a moderate overdensity near GN-z11 \citep{Tacchella2023GNz11, Scholtz2024A&A}, whose statistical significance has yet to be firmly established.

We report the discovery of a prominent overdensity candidate at $z\approx10.5$. It contains 18 galaxies with consistent photometric redshifts within a radius of 3\,arcmin. For comparison, the candidate proto-cluster near GN-z11 only contains six objects with comparable brightness. The number of member galaxies in this overdensity even exceeds most known overdensities at $z\approx7$ \citep[e.g.,][]{Helton2024overdense}. We examine the impact of this overdense environment on galaxy properties and discuss the implications for cosmic reionization. This study adopts a $\Lambda$CDM cosmology with $H_0 = 68\,\mathrm{ km\, s^{-1} \,Mpc^{-1}}$, $\Omega_m = 0.31$, and $\Omega_\Lambda = 0.69$ according to the full-mission Planck measurements (\hspace{-3pt}\citealt{Planck2020CMB}).  In this cosmology, 1 arcsec corresponds to 4 proper kpc (pkpc) and 48 comoving kpc (ckpc) at $z=10.5$.

\begin{figure*}
    \centering
    \includegraphics[width=0.7\linewidth]{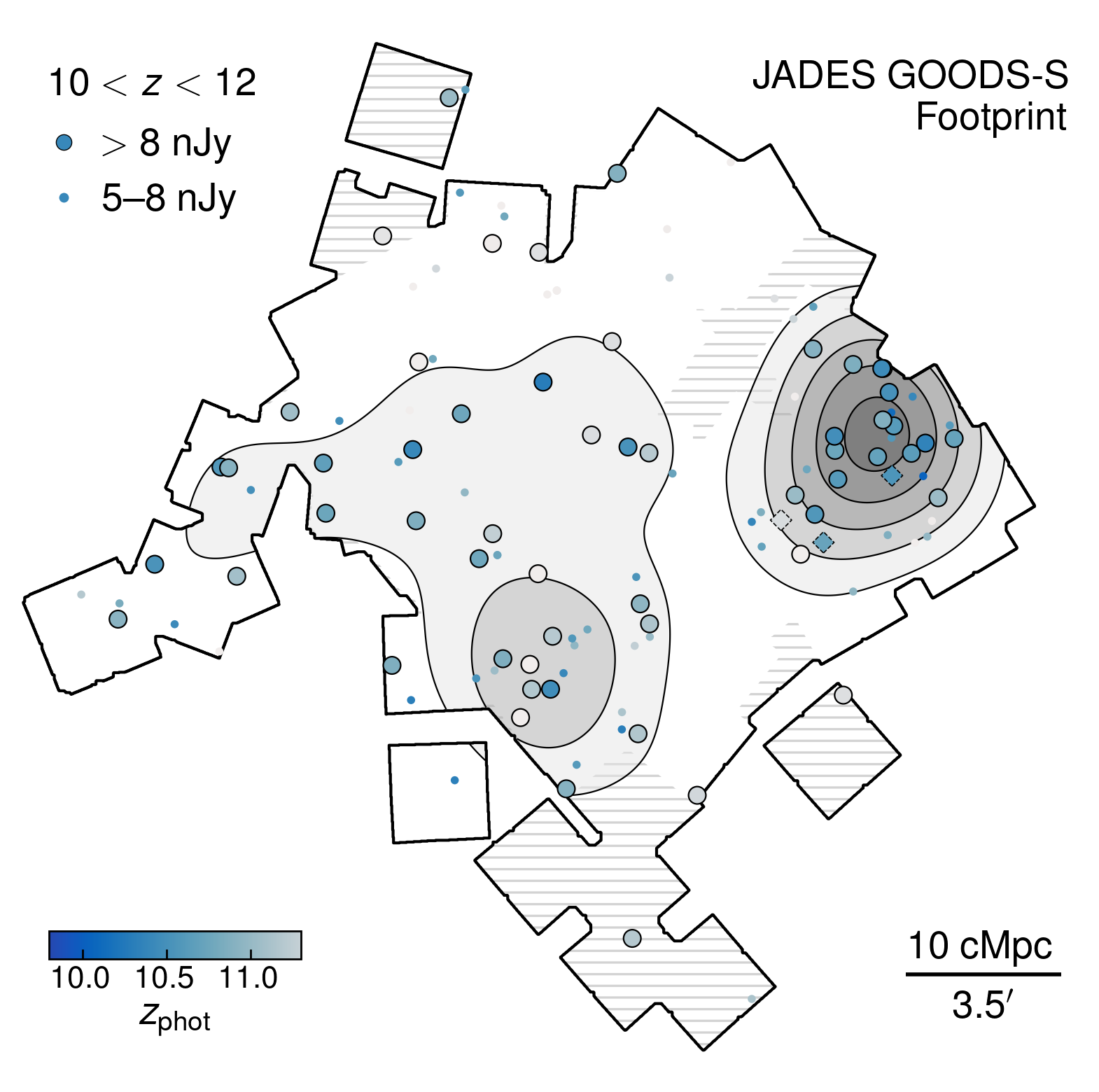}
    \caption{Spatial distribution of galaxy candidates at $10 < z_\mathrm{phot} < 12$ from the JADES photometric sample, color-coded by photometric redshift derived with the {\tt EAZY} package. Large circles mark bright sources with F356W fluxes above 8\,nJy (29.1\,AB\,mag); small circles mark sources between 5 and 8\,nJy. The bold black polygon delineates the JADES/NIRCam GOODS--S footprint in F115W. All areas within the footprint, except hatched regions, have a $7\sigma$ point-source depth deeper than 8\,nJy in F115W \citep{Johnson2026}. Contours show the estimated density field at the mean density and 1$\sigma$, 2$\sigma$, 3$\sigma$, and 4$\sigma$ above the mean density of bright sources, derived with a kernel density estimator following \cite{Helton2024overdense}. The overdensity is on the west side of the field, with a peak galaxy number density exceeding four times the mean. Three less secure objects within the overdensity, shown as diamonds with dashed outlines, are excluded from the density estimate and further analysis and are discussed in Appendix~\ref{sec:dubious}. North is up, and east is to the left.}
    \label{fig:overdensity}
\end{figure*}

\section{The Overdensity} \label{sec:overdensity}
 Deep imaging observations were obtained primarily through the JADES Guaranteed Time Observations (GTO) programs \citep{Eisenstein2023JADES} and the JADES Original Field (JOF) program 3215 \citep{Eisenstein2025ApJS}, with additional data from JWST programs 2514, 3990, 5997, and 6434 (\citealt{Williams2025PANORAMIC,Morishita2025survey, Sun2025SAPPHIRES}; Looser et al. in prep.). The overdensity resides in the JADES medium-depth field in GOODS--S, covered by 9--14 NIRCam bands, with a typical exposure time of 6~hours per band. Our analysis uses the latest JADES DR5 NIRCam imaging products \citep{Johnson2026} and the NIRCam source catalogs \citep{Robertson2026}. The 7$\sigma$ point-source depth in F115W in this region is $\sim8$\,nJy \citep{Johnson2026, Robertson2026}, similar to that achieved across most of the JADES medium-depth field. The deep F115W imaging provides a robust selection of galaxies at $z\gtrsim10$ through Ly$\alpha$ dropouts. 

We identify the overdensity from a photometric redshift catalog of $z>8$ candidates in a companion paper \citep{Hainline2026}. They estimate photometric redshifts using the {\tt EAZY} software \citep{Brammer2008EAZY} based on source identification and photometry from the JADES DR5 NIRCam source catalog \citep{Robertson2026}. To ensure robust photometric redshifts, \cite{Hainline2026} further require the sources to have a signal-to-noise ratio (SNR) $>$ 5 in at least two photometric bands, and a $\Delta\chi^2>4$ between the $z_\mathrm{phot}>10$ solution and any $z < 7$ solution. We refer to \cite{Hainline2026} for details of the photometric redshift selection. Figure~\ref{fig:overdensity} shows the spatial distribution of sources at $10<z_\mathrm{phot}<12$ that are brighter than $8$\,nJy in F356W, where the fluxes are taken from the JADES {\tt CIRC1} measurements (0.1$''$-radius apertures). We restrict attention to these bright sources because they have high signal-to-noise detections and more secure redshift constraints. Compared to shorter-wavelength bands, F356W traces star formation on longer timescales \citep{Conroy2013ARA&A} and is less sensitive to dust attenuation and metal emission lines. Three sources near the overdensity, marked as blue circles with dashed edges in Figure~\ref{fig:overdensity}, have $\sim2.5\sigma$ detections in F115W and thus could be at $z<10$. Given the ambiguity of this SNR regime, we exclude these objects from the overdensity sample and all subsequent analyses. Their properties are discussed in Appendix~\ref{sec:dubious}.

The overdensity appears as a concentration of galaxies on the west side of the JADES GOODS--S field  (Figure~\ref{fig:overdensity}). Within the mean-density contour, it contains 18 galaxies with clear Ly$\alpha$ dropout signatures and consistent photometric redshifts near $z_\mathrm{phot}=10.5$. Given its proximity to the edge of the footprint, the overdensity may have more members beyond the observed area.  The overdensity is not an artifact of selection bias since this area has depth and band coverage similar to the majority of the JADES field (\citealt{Johnson2026}). Except for a few hatched areas in Figure \ref{fig:overdensity}, the JADES GOODS--S field has a depth of 8\,nJy at $>7\sigma$ significance in F115W, ensuring robust Ly$\alpha$ dropout selection for sources brighter than 8\,nJy across the entire field. Neither the photometric-redshift criteria nor the flux threshold preferentially increase the number of detections in the overdensity region.

The overdensity far exceeds expectations from a random Poisson distribution. The region contains 18 galaxies within an area of $\sim15\,\mathrm{arcmin}^2$, whereas the entire GOODS--S field spans $209\,\mathrm{arcmin}^2$ and has only 56 objects under the same selection, including those in the overdensity. Therefore, the significance of the overdensity is $\delta_{\rm gal} = \rho_{\rm gal}/\langle \rho_{\rm gal}\rangle - 1 = 3.5$, implying a galaxy surface density $\sim4$ times higher than the field average. Comparing with the luminosity function derived from multiple programs in \cite{McLeod2024MNRAS}, we find a similar overdensity significance of $\delta_{\rm gal} = 3.6$, for the photometric redshift interval $\Delta z_\mathrm{phot}\sim1$ within the overdensity. Assuming Poisson statistics for galaxy counts, this corresponds to a $7\sigma$ deviation from the mean, where the standard deviation is $\sigma = \sqrt{\langle \rho_{\rm gal}\rangle}$.  This highly significant excess indicates a prominent overdensity that is unlikely to arise from chance alignment. It also implies a contamination rate of $\sim20\%$ (four galaxies). The low contamination suggests that statistical inferences based on this sample are robust.

The overdensity is unlikely to be mimicked by a concentration of Balmer-break galaxies at $z=2$--3. First, it is extremely rare for stellar Balmer breaks to produce such significant dropout signatures. The median flux ratio of F115W($<84\%$)/F150W is 0.051, considering the Bayesian 84\% upper limit of the F115W flux assuming a non-negative uniform prior (Eisenstein et al. in prep.). Such flux ratios far exceed normal Balmer breaks \citep{Mintz2026Balmer}. Second, we find no evidence for clustering of $z=2$–3 galaxies in this region. Third, given their faintness, many galaxies would likely be satellites at low redshift, especially in cluster environments. We therefore search for potential central galaxies within $3''$ of each source, but the number of bright neighbors (F356W $>$ 20\,nJy) is consistent with a random distribution. These considerations further reinforce the interpretation that the overdensity lies at high redshifts.

\begin{figure*}
    \centering
    \includegraphics[width=0.99\textwidth]{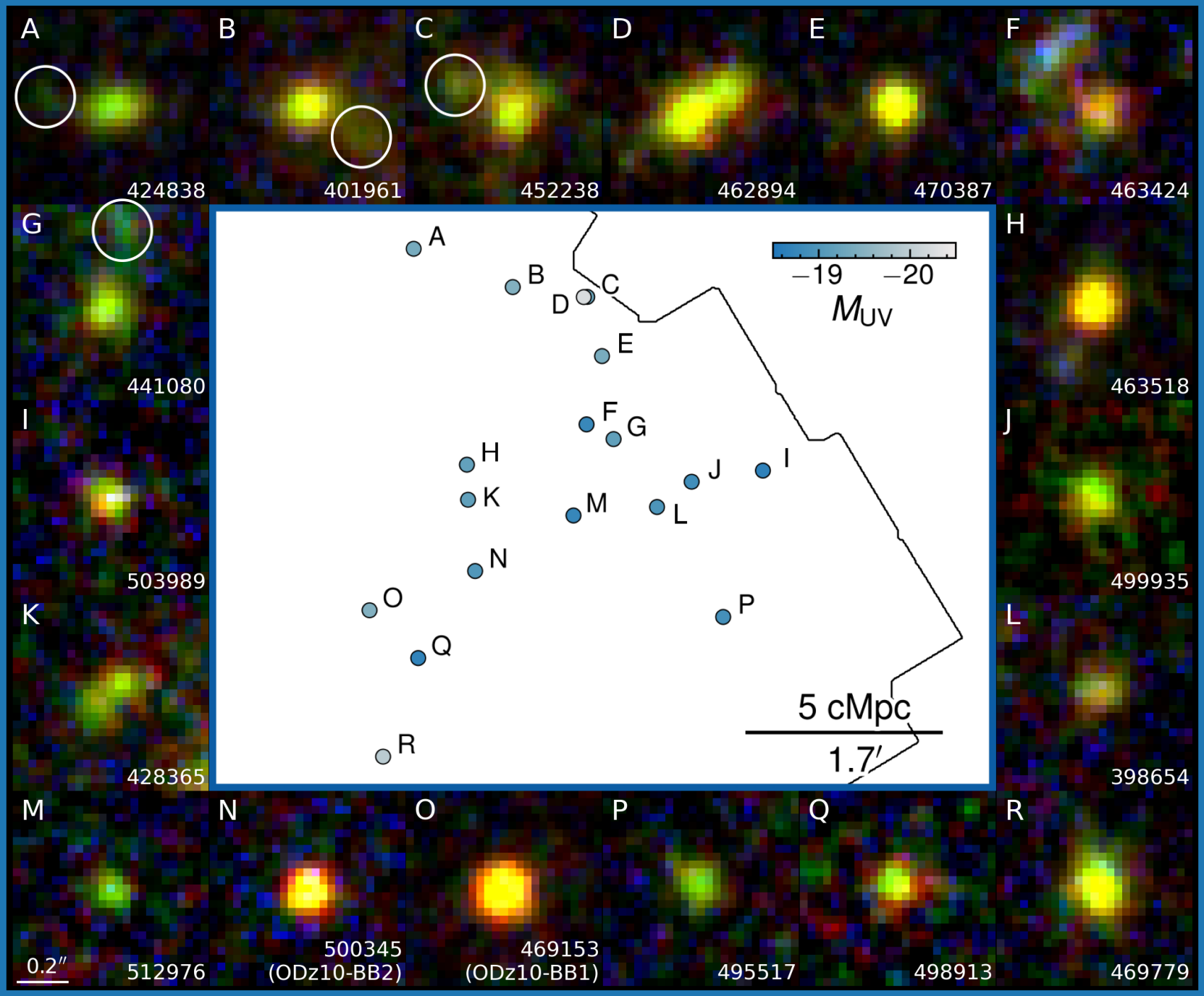}
    \caption{Composite view of the $z\approx10.5$ overdensity.  The central panel shows the relative projected positions of member galaxies, color-coded by rest-frame UV absolute magnitude $M_{\rm UV}$. The surrounding panels present JWST/NIRCam cutouts, using the F115W, F277W, and F444W bands as blue, green, and red, respectively, with F115W as the dropout band. Each cutout has a size of  $0.75^{\prime\prime}\times0.75^{\prime\prime}$; the scale bar indicates $0.2^{\prime\prime}$ (0.8\,pkpc). The upper left of each cutout shows object labels ordered by declination, in correspondence with the central panel, and the lower right shows their NIRCam IDs. Objects with close companions that are also F115W dropouts are marked by circles. Objects D and K (IDs 462894 and 428363) each consist of two components at consistent photometric redshifts. The blue companion of object F (ID 463424) is at $z_\mathrm{phot}=1.86$ and not associated with the object.}
\end{figure*}

\section{Galaxy Properties} \label{sec:galaxy}
\subsection{Measurement}
\label{sec:measure}

We perform photometry and structural decomposition using the {\tt ForcePho} package (B. D. Johnson in prep.; see also \citealt{Robertson2023NatAs, Tacchella2023GNz11, Baker2025ForcePho}). {\tt ForcePho} models the targets and their subcomponents each with a \sersic\ profile and samples model parameters using the Hamiltonian Monte Carlo Markov Chain (MCMC) method. It uses individual exposure images with dithering to achieve subpixel resolution, providing robust size measurement of even barely resolved objects. We simultaneously fit multiband NIRCam images with a common \sersic\ model convolved with corresponding point spread functions (PSFs) from STPSF (Version 2.0.0; \citealt{Perrin2014WebbPSF}). For computational efficiency, {\tt ForcePho} approximates the PSF as a mixture of six Gaussians, a simplification whose impact has been found to be negligible for faint sources \citep{Baker2025ForcePho}. We adopt flat priors for the \sersic\ indices from 0.2 to 8 and for the half-light radii from 1\,mas to 1\,arcsec. For galaxies with substructures, we visually identify individual components and fit them with separate \sersic\ models. Results of the photometry are presented in Table~\ref{tab:phot}. 

We perform spectral energy distribution (SED) fitting on the multiband photometry with the {\tt Prospector} software \citep{Johnson2021ApJS}. Redshifts are treated as free parameters, and we consider IGM absorption according to \cite{Madau1995ApJ}. We do not include damped Ly$\alpha$ absorption (DLA) because it is unconstrained by the photometric data and not well understood in overdense environments. Considering typical DLA effects may lower the inferred redshift by $\sim0.2$ \citep{Hainline2026}. We model the stellar populations using Flexible Stellar Population Synthesis (FSPS; \citealt{Conroy2009ApJ}) with Mesa Isochrones and Stellar Tracks (MIST) isochrones \citep{Choi2016ApJ} assuming a \cite{Chabrier2003} initial mass function in the $0.08$--120\,$M_\odot$ mass range. We adopt a ten-bin non-parametric star formation history (SFH) with the first bin covering a look-back time of 0--5\,Myr and the remaining bins logarithmetically spaced up to $z = 20$. We adopt a simulation-motivated rising SFH prior following \cite{Wu2025ApJ} to account for the expected rising SFHs at high redshifts \citep{Turner2025risingSFH}. We model birth-cloud dust attenuation associated with young stars following \cite{Tacchella2022} and galaxy-scale dust with flexible attenuation curves \citep{Kriek2013dust}. Nebular emission is computed self-consistently with the photoionization code {\tt CLOUDY} \citep{Byler17}. By default of {\tt Prospector}, the ``frac\_obrun'' parameter \citep{Byler17} is set to zero to prevent overfitting. We assume flat priors for gas-phase metallicity and ionization parameter in log space.  To account for uncertainties in Ly$\alpha$ radiative transfer and IGM transmission, we allow the  Ly$\alpha$ line intensity to vary via a free multiplicative factor between 0 and 1.5 relative to the {\tt CLOUDY} prediction. The fitting results are summarized in Table~\ref{tab:gal_props}.  We further estimate the UV slope $\beta$,  defined as the power index in $F_\lambda\propto\lambda^\beta$, from the {\tt Prospector} posterior spectra over the rest-frame wavelength range 1340--2600\,\AA \ in the spectral windows defined by \citet{Calzetti1994window}. We infer the absolute UV magnitude from the posterior spectra at rest-frame 1500\,\AA, which include both photometry and redshift uncertainties. Our estimates differ slightly from those of \citet{Hainline2026} because our photometry includes extended emission from the galaxies and their subcomponents, and because we adopt a different SED modeling framework. The photometric redshifts inferred with {\tt Prospector} are slightly lower than {\tt EAZY} results due to the difference in Ly$\alpha$ modeling.

\begin{deluxetable*}{cccccccccc}
\tablecaption{Properties of member galaxies in the $z\approx10.5$ overdensity \label{tab:gal_props}}
\tablehead{
\colhead{ID} &
\colhead{Component} &
\colhead{RA} &
\colhead{Dec} &
\colhead{$M_{\rm UV}$} &
\colhead{$z_{\rm phot}$} &
\colhead{$\beta$} &
\colhead{$r_{\rm half}$} &
\colhead{$\log M_\star$} &
\colhead{SFR$_{30}$} \\
\colhead{} &
\colhead{} &
\colhead{(deg)} &
\colhead{(deg)} &
\colhead{(mag)} &
\colhead{} &
\colhead{} &
\colhead{(pc)} &
\colhead{$(M_\odot)$} &
\colhead{$(M_\odot\,\mathrm{yr}^{-1})$} 
}
\startdata
398654 & \nodata & 52.955667 & $-27.808922$ & $-19.0^{+0.2}_{-0.3}$ & $9.3 ^{+1.1}_{-0.6}$ & $-2.3 ^{+0.1}_{-0.1}$ & $188 ^{+41}_{-88}$ & $7.9 ^{+0.6}_{-0.5}$ & $2.0 ^{+3.9}_{-1.1}$ \\
401961 & \nodata & 52.980418 & $-27.775546$ & $-19.5^{+0.1}_{-0.1}$ & $9.7^{+0.5}_{-0.3}$ & $-2.2^{+0.1}_{-0.1}$ & $240^{+51}_{-51}$& $8.6^{+0.4}_{-0.3}$ & $8.0^{+4.6}_{-4.3}$\\
424838 & \nodata & 52.997421 & $-27.769738$ & $-19.4^{+0.1}_{-0.1}$ & $11.0 ^{+0.6}_{-0.6}$ & $-2.4 ^{+0.2}_{-0.2}$ & $187 ^{+12}_{-12}$ & $7.9 ^{+0.4}_{-0.5}$ & $2.2 ^{+3.0}_{-1.2}$ \\
428365 & total & 52.988090 & $-27.807818$ & $-19.2^{+0.1}_{-0.2}$ & $10.1 ^{+0.8}_{-0.5}$ & $-2.3 ^{+0.1}_{-0.1}$ & \nodata & $8.2 ^{+0.3}_{-0.4}$ & $3.6 ^{+1.7}_{-2.3}$ \\
\nodata & comp1 & 52.988113 & $-27.807835$ & $-18.9^{+0.2}_{-0.3}$ & $10.8 ^{+1.3}_{-0.7}$ & $-2.4 ^{+0.3}_{-0.3}$ & $117 ^{+37}_{-31}$ & $7.8 ^{+0.5}_{-0.5}$ & $1.3 ^{+1.7}_{-0.7}$ \\
\nodata & comp2 & 52.988082 & $-27.807808$ & $-18.2^{+0.2}_{-0.3}$ & $10.9 ^{+1.2}_{-0.9}$ & $-2.4 ^{+0.2}_{-0.2}$ & $96 ^{+31}_{-31}$ & $7.3 ^{+0.3}_{-0.3}$ & $0.7 ^{+0.5}_{-0.3}$ \\
441080 & \nodata & 52.963122 & $-27.798615$ & $-19.2^{+0.1}_{-0.1}$ & $10.3 ^{+0.5}_{-0.4}$ & $-2.6 ^{+0.2}_{-0.2}$ & $96 ^{+19}_{-19}$ & $8.1 ^{+0.2}_{-0.3}$ & $3.5 ^{+1.1}_{-1.2}$ \\
452238 & \nodata & 52.967645 & $-27.777034$ & $-19.3^{+0.1}_{-0.1}$ & $10.5 ^{+0.7}_{-0.4}$ & $-1.9 ^{+0.2}_{-0.2}$ & $140 ^{+25}_{-25}$ & $8.5 ^{+0.3}_{-0.5}$ & $8.5 ^{+5.4}_{-5.1}$ \\
462894 & total & 52.968239 & $-27.777062$ & $-20.2^{+0.1}_{-0.2}$ & $10.5 ^{+0.9}_{-0.5}$ & $-2.2 ^{+0.1}_{-0.1}$ & \nodata & $8.6 ^{+0.4}_{-0.5}$ & $10.6 ^{+7.5}_{-6.3}$ \\
\nodata & comp1 & 52.968249 & $-27.777068$ & $-19.9^{+0.1}_{-0.2}$ & $10.4 ^{+0.7}_{-0.5}$ & $-2.3 ^{+0.3}_{-0.3}$ & $257 ^{+38}_{-19}$ & $8.4 ^{+0.3}_{-0.4}$ & $7.9 ^{+4.5}_{-4.3}$ \\
\nodata & comp2 & 52.968202 & $-27.777041$ & $-19.0^{+0.1}_{-0.1}$ & $10.4 ^{+0.6}_{-0.4}$ & $-2.3 ^{+0.1}_{-0.1}$ & $153 ^{+25}_{-13}$ & $8.3 ^{+0.3}_{-0.4}$ & $3.4 ^{+1.6}_{-2.1}$ \\
463424 & \nodata & 52.967760 & $-27.796385$ & $-18.8^{+0.3}_{-0.2}$ & $11.2 ^{+0.8}_{-1.3}$ & $-2.1 ^{+0.1}_{-0.1}$ & $149 ^{+24}_{-24}$ & $7.6 ^{+0.9}_{-0.2}$ & $1.1 ^{+0.6}_{-0.4}$ \\
463518 & \nodata & 52.988288 & $-27.802503$ & $-19.2^{+0.1}_{-0.1}$ & $11.7 ^{+0.3}_{-0.3}$ & $-1.5 ^{+0.3}_{-0.3}$ & $91 ^{+6}_{-6}$ & $8.2 ^{+0.2}_{-0.2}$ & $5.0 ^{+2.2}_{-1.0}$ \\
469153 & \nodata & 53.004990 & $-27.824597$ & $-19.5^{+0.1}_{-0.1}$ & $10.1 ^{+0.4}_{-0.4}$ & $-2.0 ^{+0.1}_{-0.1}$ & $76 ^{+7}_{-7}$ & $9.2 ^{+0.1}_{-0.1}$ & $0.1 ^{+1.2}_{-0.1}$ \\
469779 & \nodata & 53.002664 & $-27.846805$ & $-19.9^{+0.1}_{-0.1}$ & $11.3 ^{+0.4}_{-0.4}$ & $-2.5 ^{+0.2}_{-0.2}$ & $160 ^{+6}_{-6}$ & $8.5 ^{+0.3}_{-0.3}$ & $7.8 ^{+3.3}_{-2.7}$ \\
470387 & \nodata & 52.965106 & $-27.786018$ & $-19.4^{+0.1}_{-0.1}$ & $10.3 ^{+0.9}_{-0.5}$ & $-2.2 ^{+0.2}_{-0.2}$ & $108 ^{+7}_{-7}$ & $8.4 ^{+0.3}_{-0.4}$ & $5.9 ^{+4.0}_{-2.9}$ \\
495517 & \nodata & 52.944313 & $-27.825610$ & $-18.9^{+0.1}_{-0.1}$ & $10.8 ^{+0.4}_{-0.4}$ & $-2.3 ^{+0.1}_{-0.1}$ & $121 ^{+13}_{-13}$ & $8.3 ^{+0.2}_{-0.3}$ & $3.8 ^{+1.8}_{-2.3}$ \\
498913 & \nodata & 52.996646 & $-27.831849$ & $-18.7^{+0.1}_{-0.1}$ & $10.3 ^{+0.5}_{-0.4}$ & $-1.7 ^{+0.1}_{-0.1}$ & $62 ^{+19}_{-19}$ & $8.7 ^{+0.3}_{-0.4}$ & $9.3 ^{+6.0}_{-6.8}$ \\
499935 & \nodata & 52.949725 & $-27.805090$ & $-18.9^{+0.1}_{-0.1}$ & $10.0 ^{+0.5}_{-0.6}$ & $-1.9 ^{+0.1}_{-0.1}$ & $166 ^{+13}_{-13}$ & $8.5 ^{+0.3}_{-0.6}$ & $5.3 ^{+4.6}_{-4.1}$ \\
500345 & \nodata & 52.986872 & $-27.818640$ & $-19.0^{+0.1}_{-0.1}$ & $9.8 ^{+0.4}_{-0.4}$ & $-2.2 ^{+0.1}_{-0.1}$ & $9 ^{+7}_{-7}$ & $8.7 ^{+0.1}_{-0.2}$ & $3.1 ^{+4.1}_{-2.5}$ \\
503989 & \nodata & 52.937501 & $-27.803392$ & $-18.7^{+0.2}_{-0.3}$ & $9.8 ^{+1.1}_{-0.8}$ & $-2.1 ^{+0.1}_{-0.1}$ & $94 ^{+7}_{-7}$ & $8.0 ^{+0.5}_{-0.5}$ & $2.1 ^{+2.2}_{-1.0}$ \\
512976 & \nodata & 52.970003 & $-27.810213$ & $-18.7^{+0.1}_{-0.1}$ & $10.5 ^{+0.4}_{-0.4}$ & $-2.4 ^{+0.2}_{-0.2}$ & $21 ^{+13}_{-13}$ & $8.0 ^{+0.2}_{-0.2}$ & $2.6 ^{+0.8}_{-0.9}$ \\
\enddata
\tablecomments{Columns are as follows: (1) NIRCam source ID in the JADES DR5 catalog \citep{Robertson2026}; (2) structural component (total or individual components identified from {\tt ForcePho} modeling); (3–4) right ascension and declination in the ICRS coordinate system; (5) absolute UV magnitude at rest-frame 1500 $\mathrm{\AA}$, $M_{\mathrm{UV}}$; (6) photometric redshift inferred from {\tt Prospector} SED fitting; (7) UV continuum slope $\beta$ measured from {\tt Prospector} posterior spectra; (8) half-light radius $r_{\mathrm{half}}$ in proper parsecs, derived from multi-band \sersic\ fitting with {\tt ForcePho}; (9) stellar mass inferred from {\tt Prospector} assuming a Chabrier IMF; (10) star formation rate averaged over the past 30\,Myr. }
\end{deluxetable*}

\subsection{Possible Balmer Break}
\label{sec:balmer_break}

\begin{figure}
    \centering
    \includegraphics[width=0.45\textwidth]{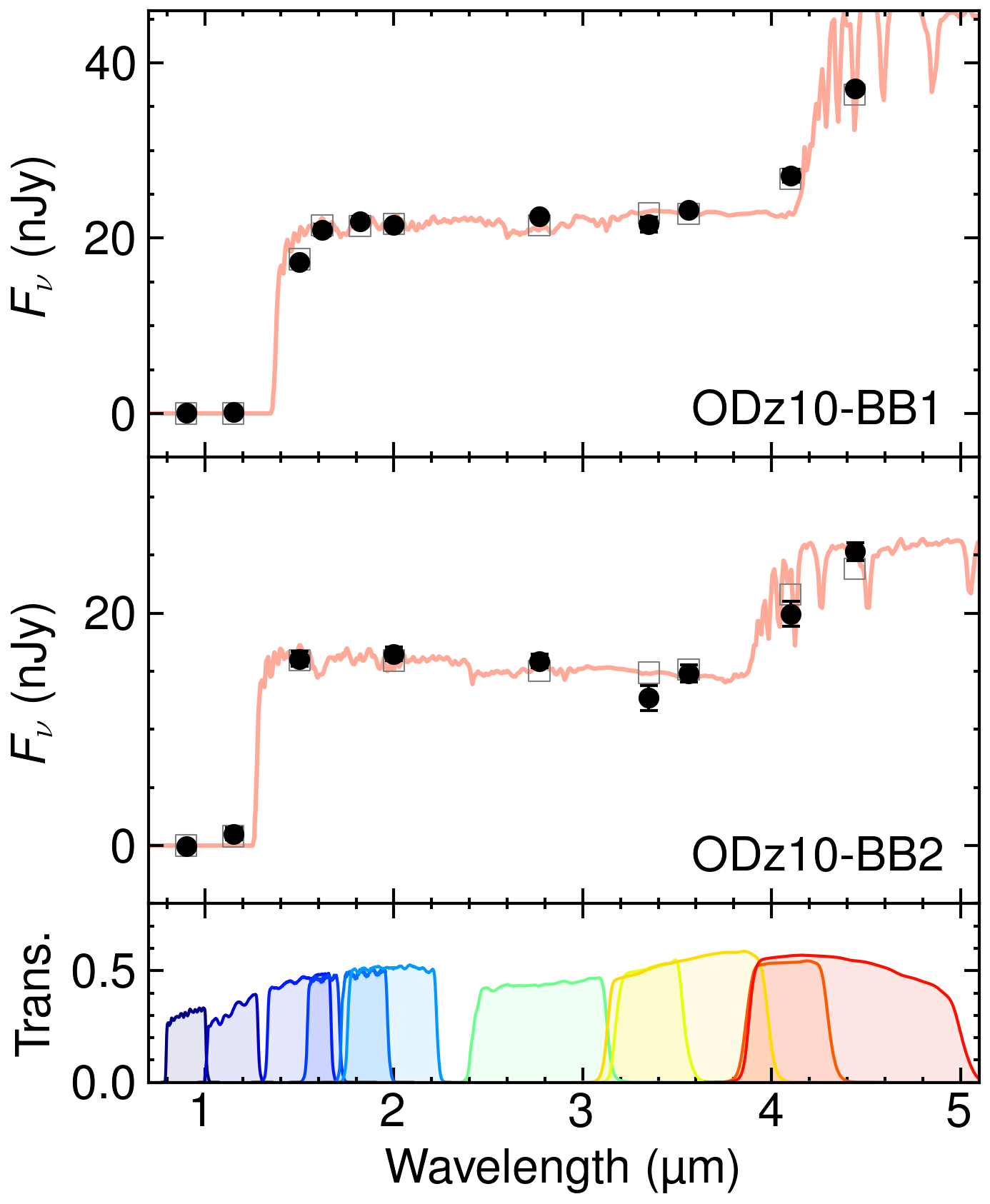}
    \caption{SED of objects with potential Balmer breaks. The upper two panels show NIRCam photometry with error bars (black dots), the best-fit galaxy spectra from {\tt Prospector} (red curves), and synthetic photometry using the best-fit spectra (gray squares). The lower panel shows the transmission curves of NIRCam filters. The Balmer break feature is unlikely to be caused by emission lines, as H$\beta$ and [\ion{O}{3}] fall outside the F444W band at $z>9$, while other lines within F444W are rarely strong enough to produce the $\sim 65\%$ flux excess.}
    \label{fig:balmer_break}
\end{figure}

Two objects in the overdensity, designated ODz10-BB1 and ODz10-BB2 (IDs 469153 and 500345), show photometric evidence of Balmer breaks (Figure~\ref{fig:balmer_break}). Their F444W fluxes exceed rest-frame UV continua by $62\%$ and $65\%$, respectively, with S/N = 28 and 13 for the excess detection, where the continua are estimated from the mean of F277W and F356W. The excesses are lower limits of the Balmer break strengths because the F444W band may include contribution from wavelengths blueward of the Balmer break.

The excesses are unlikely to be caused by emission lines. At $z>9$, H$\beta$ and [\ion{O}{3}]$\lambda$4959,\,5007 lines fall outside the F444W band and therefore cannot contribute. The overdensity is securely at $z>9$, as indicated by the clear Ly$\alpha$ dropout signatures in our deep F115W imaging. The remaining emission lines within F444W, primarily [\ion{O}{2}], [\ion{Ne}{3}], and Balmer lines, are rarely strong enough to produce the $\sim 65\%$ flux excess, which requires a rest-frame equivalent width of $\sim550\,\mathrm{\AA}$. For comparison, the combined equivalent width of all lines within F444W in GN-z11 is only 219\,$\mathrm{\AA}$ \citep{Bunker2023GNZ11}.  That said, a $z\lesssim9$ interpretation cannot be entirely excluded. In principle, extreme DLA absorption, requiring neutral-hydrogen column densities of $N_\mathrm{H\,I}>10^{23}\,\mathrm{cm^{-2}}$, could suppress the F115W flux below our detection limit, although such cases are rare \citep{Mason2026DLA}. In this scenario, the two objects would be  lower-redshift extreme-DLA interloper projected onto the $z\approx10.5$ overdensity, rather than members of it---the overdensity should be at $z\approx10.5$ because it is implausible for all 18 member galaxies to have such extreme column densities simultaneously. Finally, these objects cannot be brown dwarfs given their flat SED in the rest-frame UV across wide and medium bands and the non-detection in F115W, distinct from the SEDs of brown dwarfs \citep{Hainline2024ApJ}. 

Such Balmer breaks, if confirmed, may arise from either the LRD phenomenon \citep[e.g.,][]{Matthee2024LRD} or from an evolved stellar population. The flux increase from F410M to F444W matches the V-shaped SED typical of LRDs \citep{Setton2025LRD}. We find their rest-frame optical continuum slopes, $\beta_{\rm optical}$ (defined as $f_\lambda \propto \lambda^\beta$), are $1.9 \pm 0.3$ and $1.1 \pm 0.7$, derived from power-law fits to the F410M and F444W photometry considering filter throughput. These slopes are characteristic of LRDs \citep{Kocevski2025LRD}. Their compact morphologies are also consistent with LRDs: the posterior half-light radii of ODz10-BB1 and ODz10-BB2 are $24\pm4$ and $3\pm2$ mas ($96 \pm 16$ and $12 \pm 8$ pc), respectively. The point-like morphology of ODz10-BB2 is consistent with known LRDs, and although ODz10-BB1 is slightly extended, it could still be an LRD with a host galaxy that contributes the extended light \citep[e.g.,][]{Chen2025ApJ, Inayoshi2025LRD, Rinaldi2025ApJ}.

An evolved stellar population with a Balmer break could also produce the rise in F410M and F444W. Figure~\ref{fig:balmer_break} shows the best-fit stellar spectra in red curves, modeling the Balmer break with an evolved stellar population. The close match of synthetic photometry (gray squares) with observations indicates that stellar Balmer breaks can explain the F410M and F444W excess. The compact morphologies are also consistent with massive quiescent galaxies at $z\sim7$ \citep{Weibel2025ApJ}, although such quiescent galaxies are very rare \citep{Baker2025quiescent}. This scenario would imply large stellar masses: our SED fitting finds $\log\,(M_*/M_\odot)=9.24\pm0.12$ and $8.70\pm0.13$ for ODz10-BB1 and ODz10-BB2, respectively. It also implies surprisingly mature stellar populations in the early Universe. We will further discuss the implications in Section~\ref{sec:discuss_galaxy}.

\subsection{Size--mass Relation}
\begin{figure*}
    \includegraphics[width=0.99\textwidth]{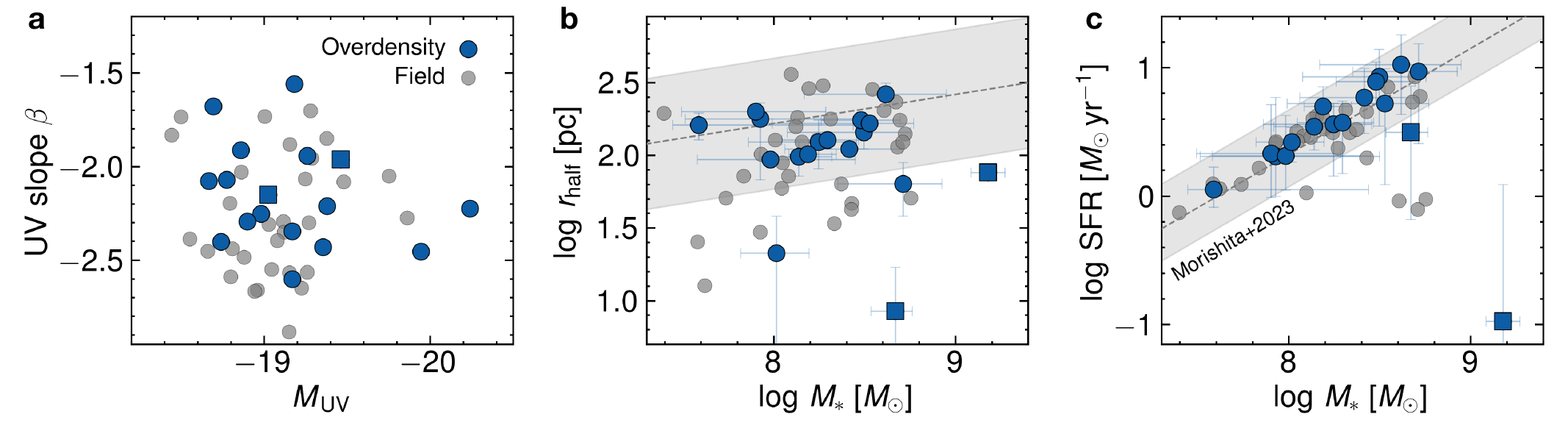}
    \caption{(a) UV continuum slope $\beta$ versus absolute UV magnitude $M_{\mathrm{UV}}$. (b) Half-light radius versus stellar mass, where sizes are measured from multiband \sersic\ fitting using {\tt ForcePho}. (c) Star formation rate averaged over the past 30\,Myr versus stellar mass. Blue points denote galaxies in the overdensity; blue squares highlight those with potential Balmer breaks.  Gray points show field galaxies at similar redshifts ($10 < z_{\mathrm{phot}} < 12$) and brightness (F356W $> 8$~nJy) in GOODS--S. Dashed lines and shaded regions indicate the size-mass relation and the star formation main sequence, along with their intrinsic scatters, for galaxies at $z\approx10$, inferred from multiple surveys  by \cite{Morishita2024size}.}
    \label{fig:size-sfr}
\end{figure*}

Figure~\ref{fig:size-sfr}b compares the size-mass distribution of galaxies inside the overdensity with field galaxies. The sizes are measured from multiband fitting with {\tt ForcePho} using \sersic\ profiles. The primary constraints come from the F150W and F200W bands, probing rest-frame 1130--2000\,$\mathrm{\AA}$, where the PSFs are the sharpest. We adopt circularized half-light radii $r_\mathrm{half}$ because the sources are marginally resolved and the circularized radii are more stable to noise than semi-major radii. For sources with multiple components, we plot the size of the primary component only.

Most galaxies in the overdensity have $r_\mathrm{half}\approx 200$\,pc, whereas three objects (441080, 512976, and ODz10-BB1) are significantly smaller and consistent with point sources. ODz10-BB1 has a potential Balmer break (Section~\ref{sec:balmer_break}), while 441080 and 512976 are typical young galaxies with blue UV slopes, similar to many other compact star-forming galaxies at high redshifts (e.g., \citealt{Witstok2025GSz13Lya, Wu2025ApJ}). Notably, 441080 has diffuse emission and a companion in the north, although its dominant central component is very compact.

We do not find a significant difference in the size--mass relation compared to field galaxies, although galaxies in the overdensity are slightly more massive. The field sample includes all objects in GOODS--S at similar photometric redshift ($10<z_\mathrm{phot}<12$) with the same brightness threshold (F356W $>8$\,nJy). We also find consistency in the size distribution with results in \cite{Morishita2024size} over our mass range, where we convert their semi-major radii to circularized radii using a median axis ratio of $b/a=0.7$. The consistency with field galaxies may suggest that the structural build-up of the galaxies has not differed significantly between dense and average environments at this early epoch.

\subsection{Star Formation Rate}

Figure~\ref{fig:size-sfr} shows the distribution in the star formation rate (SFR)--stellar mass plane. SFRs are inferred from {\tt Prospector} fits to the NIRCam SEDs, where we report SFRs averaged over the past 30\,Myr, as the rest-frame UV continuum probes SFRs on timescales of 10--50\,Myr \citep{Kennicutt2012ARA&A}. Most galaxies in the overdensity have SFR $\sim 3\,M_\odot\,\mathrm{yr^{-1}}$ and specific SFR (sSFR) $\sim 40\,\mathrm{Gyr^{-1}}$, implying stellar-mass doubling times of $\sim25$\,Myr if the SFR is sustained.

Galaxies in the overdensity have a median SFR that is $\gtrsim30\%$ higher than that of the field galaxies at $10<z_\mathrm{phot}<12$. The enhanced SFR is also reflected in their systematically higher luminosities in Figure~\ref{fig:size-sfr}a. For a fair comparison, the field sample shown in Figure~\ref{fig:overdensity} is selected with the same criteria (F356W $>$ 8\,nJy) and measured consistently with {\tt Prospector}.  While studies at $z\approx5$–7 find lower SFRs in overdensities relative to the field \citep{Helton2024ApJEvolved, Li2025metal}, our result is consistent with their inference that at earlier epochs, these galaxies should have higher SFRs than field galaxies according to their star formation histories and larger stellar masses \citep{Helton2024ApJEvolved}.

The overdensity has a total SFR of  $98\,M_\odot\,\mathrm{yr^{-1}}$ and an SFR density of $0.05\,M_\odot\,\mathrm{yr^{-1}\,cMpc^{-3}}$, where the latter assumes a line-of-sight extent comparable to the transverse scale. Both values should be regarded as lower limits, as our selection does not include faint overdensity members. They are slightly higher than, but broadly consistent with, predictions from semi-analytical models of cosmic star formation \citep[e.g.,][]{Chiang2017ApJ}. 

For comparison, the total SFR of field galaxies at $10<z_\mathrm{phot}<12$ in GOODS--S is $105\,M_\odot\,\mathrm{yr^{-1}}$, implying that the overdensity accounts for $\sim48\%$ of the total star formation. Simulations predict that proto-clusters account for 50\% of cosmic star formation at $z\approx10$ \citep{Chiang2017ApJ}, which is comparable to the fraction inferred for this overdensity. However, this estimate should be interpreted with caution, as our photometric selection may include $\sim20\%$ contamination from field galaxies projected into the overdensity (Section~\ref{sec:overdensity}), while additional members may lie outside the current observed footprint, and our analysis is restricted to relatively bright objects. Cosmic variance in GOODS--S may also affect the inferred fractions.

Finally, we consider how the choice of SFR estimator may affect these results. When estimating SFR from the UV luminosity using the \cite{Kennicutt1998} relation, after converting to the \cite{Chabrier2003} IMF following \cite{ZhangYechi2026}, the inferred $\mathrm{SFR}_{\rm UV}$ are generally lower by $\sim40\%$ than the $\mathrm{SFR}_{30}$ values from {\tt Prospector}, although the two remain consistent within the uncertainties. This offset is expected as these galaxies typically have rising star formation histories. The \citet{Kennicutt1998} UV calibration assumes continuous star formation over recent $\sim10$--100\,Myr timescales, whereas $\mathrm{SFR}_{30}$ is the mean over the past 30 Myr. Additional differences may arise from the treatment of dust attenuation and from the stellar-population and nebular-emission assumptions entering the two methods. In any case, the fractional contribution of the overdensity to the cosmic star formation at $10<z_{\mathrm{phot}}<12$ remains essentially unchanged.

\subsection{Galaxy Merger and Clumps}
Six galaxies in the overdensity show subcomponents within 1\,pkpc that are also F115W dropouts, indicating consistent redshifts. Their NIRCam IDs are 401961, 424838, 428365, 441080, 452238, and 462894. These subcomponents are likely gravitationally bound, as the expected virial radii of their halos are $\sim 8$\,pkpc, given the stellar-halo mass relation \citep{Stefanon2021halomass}.

The subcomponents may originate from galaxy mergers or clumpy star formation. However, the clumpy explanation is less likely because many of these components lie at $\sim 1$\,pkpc from the primary component, well beyond the effective radius of the host galaxy. Such offsets are difficult to explain with in-situ stellar clumps. 

The fraction of galaxies with subcomponents exceeds that of field galaxies. This fraction is 33\% inside the overdensity but is only $\sim 15\%$ among field galaxies based on our visual inspection. For comparison, quantitative analysis of \cite{Hainline2026} find 22\% galaxies at $z_\mathrm{phot}=10$--11 show extended emission, which is consistent with our estimate.

The galaxies with subcomponents also show a spatial preference inside the overdensity: five of the six galaxies are located in the northern part of the overdensity. Notably, 452238 and 462894 are separated by only 8.4\,pkpc (2.1$''$). The distance is comparable to the virial radius of their halos and thus may suggest ongoing major mergers. Follow-up spectroscopic observations will be important to confirm their physical association.

\subsection{Ly$\alpha$ Emission Estimate}

\begin{figure*}
\centering
    \includegraphics[width=0.85\textwidth]{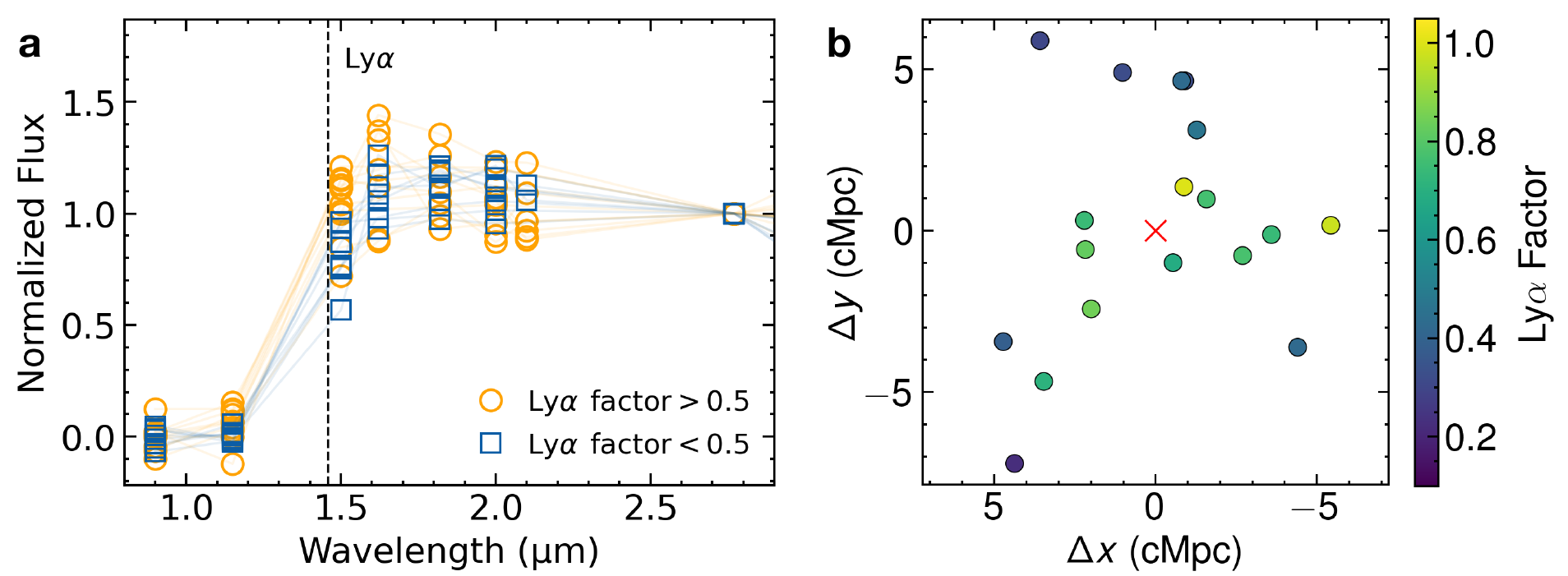}
    \caption{
(a) Galaxy SEDs normalized by the F277W flux. Orange circles denote galaxies with inferred Ly$\alpha$ factors greater than 0.5, and blue squares denote those with Ly$\alpha$ factors below 0.5. The dashed line indicates the expected Ly$\alpha$ wavelength at $z = 10.5$. (b) Projected galaxy positions color-coded by Ly$\alpha$ factor; the red cross marks the geometric center of the overdensity. We define the Ly$\alpha$ factor, an indicator of Ly$\alpha$ transmission, as the ratio of the inferred Ly$\alpha$ flux to the Ly$\alpha$ flux predicted by the nebular model. The Ly$\alpha$ factors are estimated from photometry using {\tt Prospector}, fixing all galaxies at $z = 10.5$. }
    \label{fig:lya}
\end{figure*}

Ideally, Ly$\alpha$ fluxes would be measured directly from spectroscopy \citep[e.g.,][]{Bunker2023GNZ11, Jones2024A&A, Jones2025MNRAS, Witstok2025GSz13Lya}. Here, we estimate the flux using our current photometry data, exploiting the fact that Ly$\alpha$ falls within the F150W filter and can boost its flux. Given the photometric uncertainty and the filter bandwidth, the F150W data correspond to a median $1\sigma$ line-flux sensitivity of $2.7\times10^{-19}\mathrm{\,erg\,s^{-1}\,cm^{-2}}$ (EW$_\mathrm{res} = 12\,\mathrm{\AA}$ for $M_\mathrm{UV}=-19$), sufficient to detect Ly$\alpha$ with strengths comparable to those of GN-z11 and JADES-GS-z13-0-LA \citep{Bunker2023GNZ11, Witstok2025GSz13Lya}. A major limitation, however, is the uncertainties in redshift. The F150W flux is highly sensitive to the precise redshift because it sets the location of the Ly$\alpha$ break, leading to a strong degeneracy between Ly$\alpha$ strength and redshift, which precludes meaningful constraints for individual objects. Nonetheless, this issue is alleviated by the fact that the galaxies lie in the same overdensity and are therefore expected to have similar redshifts. If their line-of-sight separations are comparable to the projected distances ($\sim10$\,cMpc), the implied redshift spread is $\Delta z \simeq 0.05$. Even allowing for large internal velocity offsets, a relative velocity of $1000\,\mathrm{km\,s^{-1}}$ only produces $\Delta z \simeq 0.04$. Both effects are negligible in the context of the F150W bandwidth. By imposing a common redshift for the galaxies, we are therefore able to infer the spatial distribution of Ly$\alpha$ fluxes.

To implement this, we sample a redshift grid from $z = 10.0$ to $11.4$ in steps of $0.1$. At each grid point, we run {\tt Prospector} SED fits for every galaxy with the redshift held fixed. As in Section~3.1, we fit a Ly$\alpha$ factor, defined as the ratio of the best-fit Ly$\alpha$ flux to the nebular Ly$\alpha$ flux predicted by the {\tt cloudy} model. The modeling includes neutral IGM absorption following \citet{Madau1995ApJ}, with a free scaling parameter that accounts for line-of-sight variations in the total opacity. We adopt a Gaussian prior on this parameter, centered at 1 with a dispersion of 0.3, as the default choice in {\tt Prospector}. The resulting posterior distribution of the Ly$\alpha$ factor has a typical $1\sigma$ width of 0.27. We do not model DLA absorption or the effect of ionized bubbles on the IGM, as they are unconstrained in the absence of spectroscopic data. They are expected to further broaden the uncertainty on the Ly$\alpha$ factor. We qualitatively discuss their potential impact later in this section.

At $z = 10.5$, the redshift favored by the joint posterior of the redshift-free modeling in Section~\ref{sec:measure}, the Ly$\alpha$ factor shows a spatial pattern: it is elevated near the center of the overdensity and decreases toward the outskirts (Figure~\ref{fig:lya}b). The radial decrease has a Spearman correlation coefficient of $-0.62$ ($p=0.003$, $n=18$). This behavior is consistent with enhanced Ly$\alpha$ transmission in the core and reduced transmission in the outskirts. 

Figure~5a provides a visual illustration of the imprint of Ly$\alpha$ emission on the observed SEDs. When the galaxies are divided by Ly$\alpha$ factor, those with higher inferred values show a  systematic relative excess in the F150W band. We further find no clear correlation between the inferred Ly$\alpha$ factor and either SFR or stellar mass, suggesting that the Ly$\alpha$ factor is not likely to be driven by effects of stellar population modeling.

The radial decline is robust across the entire redshift range. For each assumed redshift, we examine the Ly$\alpha$ factor as a function of distance from the overdensity center, where the center is defined as the arithmetic mean of the galaxy coordinates. Although the Ly$\alpha$ factors of individual galaxies vary with redshift, the collective trend is unchanged. Even at $z = 11$, the profile is similar to the $z = 10.5$ case. Fitting an exponential profile to the radial variation yields a scale radius of $6\,\mathrm{cMpc}$ at $z = 10.5$, and this value remains nearly constant between $z = 10.5$ and $11$. 

However, the photometric inference is not definitive. First, we assumed all galaxies in the overdensity share the same redshift. Nonetheless, galaxies close in projection could still be widely separated along the line of sight. A configuration in which lower-redshift galaxies lie near the projected center and higher-redshift galaxies on the edge  could reproduce the observed pattern. Second, inferring transmission requires an estimate of the intrinsic Ly$\alpha$ flux, which depends on the nebular emission model and the escape fraction of ionizing photons. If galaxies at the edge have higher escape fractions, they would produce lower Ly$\alpha$ flux even under identical IGM transmission. Finally, spatial variations of DLA and IGM absorption may also affect F150W. This effect is complicated: DLA absorption is expected to be enhanced toward the densest regions of the overdensity due to the higher abundance of massive halos along the line of sight, whereas IGM absorption may be enhanced by higher recombination rates or reduced by larger ionized bubbles, depending on the local balance between density and ionizing sources. Direct Ly$\alpha$ and redshift measurements from spectroscopy are therefore essential to test the bubble interpretation.

Finally, we note that the bubble radius inferred from photometric Ly$\alpha$ transmission is consistent with a first-order estimate of the bubble size based on the ionizing photon budget \citep{Mason2020MNRAS}:
\begin{equation}
R_{\rm ion} \simeq
\left(
\frac{3\, f_{\rm esc}\, \dot N_{\rm ion}\, t_\mathrm{age}}
{4\pi \,\bar n_{\rm H}(z)}
\right)^{1/3} \gtrsim 4\,\mathrm{cMpc}
\end{equation}
where we assume an escape fraction $f_{\rm esc}=5\%$ following \cite{Witstok2024A&A} and a characteristic duration of ionizing photon production $t_\mathrm{age}=30\,$Myr according to the median stellar age from our SED fitting. At the level of this order-of-magnitude estimate, the resulting bubble size depends only weakly on these parameters. We calculate $\bar n_{\rm H}$ according to the mean cosmological hydrogen number density at redshift $z$. The total ionizing photon production rate $\dot N_{\rm ion}$ is estimated by summing the contributions from all galaxies within the overdensity, inferred from their $M_{\rm UV}$ and $\beta$ using the prescription of \citet{Mason2020MNRAS}. This estimate should be regarded as a lower limit, as our sample includes only relatively bright galaxies, while faint sources are expected to contribute additional ionizing photons.

\section{Discussion}\label{sec:discuss}

\subsection{Implications for Early Galaxy Evolution}
\label{sec:discuss_galaxy}

The first- and second-most luminous galaxies in the overdensity are 462894 and 469779, respectively, with absolute UV magnitudes of $M_\mathrm{UV}=-20.2$ and $-19.9$. They are the brightest galaxies across the entire GOODS--S field at $10<z_\mathrm{phot}<12$. The concentration of the most luminous objects within such a small region indicates that the overdensity hosts a disproportionately large number of the brightest galaxies, providing evidence that overdensities host the most massive halos.

The UV luminosities of the brightest galaxies in the overdensity are comparable to, though slightly fainter than, those of GNz11, JADES-GS-z14-0, and GHZ2 \citep{Bunker2023GNZ11, carniani_2024, Castellano2024GHZ2}. However, none of these galaxies appears to reside in a comparably significant overdensity. While GN-z11 may reside in a candidate proto-cluster, its significance is substantially weaker than the overdensity reported here \citep{Tacchella2023GNz11, Scholtz2024A&A}. This contrasts with the expectation that the most massive halos should occupy the most overdense environments. One possible interpretation is that such galaxies can appear over-luminous even in less massive halos, a natural consequence of bursty star formation \citep[e.g.,][]{Sun2023a}.  Alternatively, another possibility is that the most massive galaxies in the overdensity are no longer the most UV luminous. Two galaxies in the overdensity show potential Balmer breaks. If these Balmer breaks arise from evolved stellar populations, the implied stellar masses of $\log (M_*/M_\odot) = 9.2$ and 8.6 are consistent with the expected descendants of over-luminous objects like JADES-GS-z14-0, given its star-formation rate. {\tt Prospector} modeling further indicates that these Balmer-break galaxies experienced a peak star-formation rate of $\sim 10\, M_\odot\,{\rm yr^{-1}}$ at $z \approx 12$, which then declined to $\sim 5\times10^{-2}\, M_\odot\,{\rm yr^{-1}}$, leaving them faint. If confirmed, these objects would provide the missing observational link to the descendants of the overluminous $z>10$ population and would imply strongly bursty star formation in the early universe \citep{Boyett2024MNRAS}. Similar massive (mini-)quenched galaxies have been identified in overdensities at $z\approx5$ \citep{Baker2026A&A}.

\subsection{Implications for Cosmic Reionization}
We find that Ly$\alpha$ transmission may be nearly transparent around the overdensity center and is consistent with declining toward its outskirts. This appears as a decrease in the F150W flux relative to the UV continuum from the center to the edge. We quantify the transmission by fitting the SEDs of all galaxies with their redshifts fixed to a common value over a redshift grid, as described in Section 3.5. The declining trend is robust because it persists across the full redshift range we tested and thus may suggest an ionized bubble forming in the overdensity.

If confirmed, the spatial variation of Ly$\alpha$ transmission would mark the earliest ionized bubble produced by a galaxy overdensity known so far. It provides a rare laboratory to study the structure of ionizing bubbles. Previous Ly$\alpha$ detections from JADES-GS-z13-1-LAE imply bubble sizes of $\sim 5$ cMpc \citep{Witstok2025GSz13Lya}, but these estimates rely on detailed radiative transfer modeling. In contrast, the overdensity enables direct constraints from the spatial distribution of Ly$\alpha$ transmission. Spectroscopic redshifts for galaxies at different line-of-sight positions would further map variations in Ly$\alpha$ transmission within the bubble \citep{Lu2025, Nikolic2025}. These measurements would place strong constraints on the physics of cosmic reionization in the early universe.

\subsection{Comparison with \textsc{Thesan} and Implications for Galaxy Formation}
We compare this galaxy population to that of the \textsc{Thesan} simulation. \textsc{Thesan} is a suite of large-volume cosmological radiation–magneto-hydrodynamic simulations of the Epoch of Reionization, combining the \textsc{IllustrisTNG} galaxy-formation framework with a self-consistent treatment of hydrogen reionization \citep{kannan_span_2022}.  However, it is well-known that current simulations underpredict the luminosity function of galaxies at $z>10$ \citep[e.g.,][]{Finkelstein2023ApJ,Donnan2024MNRAS,Finkelstein2024ApJ, Whitler2025ApJ, Weibel2026ApJ}. 

We do not find a comparable overdensity in the \textsc{Thesan} simulation. We search for overdensities in the fiducial \textsc{Thesan}-1 run, which has a box size of $95.5\,\mathrm{cMpc}$, equivalent to five times the JADES GOODS--S area. We select galaxies using the same flux limit of F356W $> 8\,\mathrm{nJy}$, corresponding to the rest-frame $U$-band in \textsc{Thesan}. Under this selection, the maximum overdensity within an $8\,\mathrm{cMpc}$ radius contains only six galaxies, far fewer than observed. The stellar masses of these galaxies range from $0.5\text{–}2.4\times10^8\,M_\odot$, with a median of $6\times10^7\,M_\odot$, which are $\sim40$\% lower than the masses of galaxies in our observed overdensity. 

However, we do recover comparable overdensities once we brighten all simulated galaxies by 0.9\,mag in an \emph{ad hoc} way. This adjustment also brings the simulated galaxy number density into agreement with JADES. In this case, the most overdense region in \textsc{Thesan} contains 17 galaxies within $5\,\mathrm{cMpc}$, of which 14 galaxies also lie within $5\,\mathrm{cMpc}$ in the $z$-direction, while the remaining three are separated by $30\text{–}60\,\mathrm{cMpc}$ and appear close only due to projection.  This overdensity generates an ionized bubble with ionized hydrogen fractions $>90\%$, computed using the median over $10\,\mathrm{cMpc}$ along the line of sight ($z$ direction). The resulting bubble has a radius of $\sim 4\,\mathrm{cMpc}$. The bubble size would likely be larger if the galaxies were truly brighter by 0.9\,mag and produced more ionizing photons. This behavior is consistent with the bubble size inferred for our observed overdensity.

Nonetheless, the overdensity is not recovered under the scenario of bursty star formation \citep[e.g.,][]{Sun2023a, Sun2023, Asada2024}. We apply a random Gaussian perturbation with a scatter of 1.05 mag to the galaxy luminosities, which  matches the  number density of JADES galaxies brighter than 8\,nJy. However, in our random realizations, the probability of producing an overdensity with at least 18 galaxies is only $6\%$, even after accounting for chance projections. This is likely because bursty star formation makes low-mass halos more likely to be observed, while low-mass halos are less clustered \citep{Mo1996}. Therefore, the existence of the overdensity may suggest that bursty star formation, though it should play a role to some extent, cannot by itself explain the full picture of early galaxy formation. Additional mechanisms that enhance the star-formation efficiency or UV luminosity while acting preferentially in massive halos, or at least without preference on low-mass halos, are required. This interpretation is consistent with the clustering analysis of \citet{Weibel2025}, which finds mild tension with bursty star formation models and instead favors an SFE that increases with halo mass.

\section{Summary and Future Prospects}
We identify an overdensity at $z_\mathrm{phot}\approx10.5$ in the JADES GOODS--S field. It contains 18 objects occupying a region only $\sim10$\,cMpc across, exceeding the average density by a factor of four. Most members are compact, low-mass, and actively star-forming, consistent with typical scaling relations. Close companions within $\sim1$\,pkpc occur more frequently than in the field, suggesting enhanced interaction and merger activity. Two compact sources show potential Balmer breaks, suggesting the presence of LRDs or unexpectedly evolved stellar populations within 500\,Myr after the Big Bang. We find tentative evidence for spatial variation in Ly$\alpha$ transmission from the photometric data, consistent with the formation of an ionizing bubble with radius $\sim6$\,cMpc.

These conclusions are necessarily provisional because they rely on photometry alone. In our upcoming JWST cycle 5 program (PI: Wu \& Eisenstein, GO 12340), we will use NIRSpec Prism and the G140M grating to obtain spectroscopic redshifts and test for the presence of Balmer breaks and Ly$\alpha$ emission. Measurements of the spatial distribution of Ly$\alpha$ flux will directly examine the presence of an ionized bubble. Finally, because Ly$\alpha$-breaks and Ly$\alpha$-lines are affected by DLA effects and resonant scattering, respectively, precise redshift measurements require non-resonant tracers. ALMA observations targeting [\ion{O}{3}]\,88\,$\mu$m emission, or JWST/MIRI LRS observations of [\ion{O}{3}] $\lambda$5007 and H$\alpha$, can provide robust redshift measurements for the brightest members, anchoring the three-dimensional structure of the overdensity. Together, these measurements would transform the photometric evidence into a definitive test of early overdense structure formation and the onset of cosmic reionization.

\begin{acknowledgments}
We thank Anna de Graaff for insightful discussions. This work is based on observations made with the NASA/ESA/CSA James Webb Space Telescope. The data were obtained from the Mikulski Archive for Space Telescopes at the Space Telescope Science Institute, which is operated by the Association of Universities for Research in Astronomy, Inc., under NASA contract NAS 5-03127 for JWST. These observations are associated with programs 1286, 1287, 2514, 3215, 3990, 5997, and 6434. This research made use of the lux supercomputer at UC Santa Cruz which is funded by NSF MRI grant AST 1828315.

DJE, BDJ, KH, JH, ZJ, BER, FS, and CNAW acknowledge support from the NIRCam Science Team contract to the University of Arizona, NAS5-02015. DJE is also supported as a Simons Investigator and by NASA through a grant from the Space Telescope Science Institute, which is operated by the Association of Universities for Research in Astronomy, Inc., under NASA contract NAS5-03127. WMB gratefully acknowledges support from DARK via the DARK fellowship and research grant (VIL54489) from VILLUM FONDEN. AJB acknowledges funding from the ``FirstGalaxies'' Advanced Grant from the European Research Council (ERC) under the European Union’s Horizon 2020 research and innovation program (Grant agreement No. 789056). AJC gratefully acknowledges support from the Cosmic Dawn Center through the DAWN Fellowship. The Cosmic Dawn Center (DAWN) is funded by the Danish National Research Foundation under grant No. 140. ECL acknowledges support of an STFC Webb Fellowship (ST/W001438/1). ALD thanks the University of Cambridge Harding Distinguished Postgraduate Scholars Programme and Technology Facilities Council (STFC) Center for Doctoral Training (CDT) in Data intensive science at the University of Cambridge (STFC grant number 2742605) for a PhD studentship. RH acknowledges funding provided by the Johns Hopkins University, Institute for Data Intensive Engineering and Science (IDIES). JMH is also supported by JWST program 3215.  TJL gratefully acknowledges support from the Swiss National Science Foundation through a SNSF Mobility Fellowship and from the NASA/JWST Program OASIS (PID 5997). RM acknowledges support by the Science and Technology Facilities Council (STFC), by the ERC through Advanced Grant 695671 ``QUENCH”, and by the UKRI Frontier Research grant RISEandFALL. RM also acknowledges funding from a research professorship from the Royal Society. BER also acknowledges support from JWST Program 3215. ST acknowledges support by the Royal Society Research Grant G125142. JAAT acknowledges support from the Simons Foundation and JWST program 3215. Support for program 3215 was provided by NASA through a grant from the Space Telescope Science Institute, which is operated by the Association of Universities for Research in Astronomy, Inc., under NASA contract NAS 5-03127. The research of CCW is supported by NOIRLab, which is managed by the Association of Universities for Research in Astronomy (AURA) under a cooperative agreement with the National Science Foundation. JW gratefully acknowledges support from the Cosmic Dawn Center through the DAWN Fellowship. The Cosmic Dawn Center (DAWN) is funded by the Danish National Research Foundation under grant No. 140.

\end{acknowledgments}

\appendix
\twocolumngrid
\restartappendixnumbering

\section{Galaxy Spectral Energy Distribution}
This appendix presents the photometry and best-fit SEDs of galaxies in the overdensity. The photometry in Table~\ref{tab:phot} is obtained using \texttt{ForcePho}, which simultaneously fits the individual galaxy components with \sersic\ profiles. Figure~\ref{fig:sed_all} shows the \texttt{Prospector} best-fit SEDs based on the total flux summed over all components.

\begin{deluxetable*}{cccc*{12}{c}}
\tabletypesize{\scriptsize}
\tablecaption{JWST/NIRCam photometry for galaxies in the overdensity.
\label{tab:phot}}
\tablehead{
\colhead{ID} & \colhead{Component} &
\colhead{F090W} & \colhead{F115W} & \colhead{F150W} & \colhead{F162M} &
\colhead{F182M} & \colhead{F200W} & \colhead{F210M} & \colhead{F250M} &
\colhead{F277W} & \colhead{F300M} & \colhead{F335M} & \colhead{F356W} &
\colhead{F410M} & \colhead{F444W} \\
\colhead{} & \colhead{} &
\colhead{(nJy)} & \colhead{(nJy)} & \colhead{(nJy)} & \colhead{(nJy)} &
\colhead{(nJy)} & \colhead{(nJy)} & \colhead{(nJy)} & \colhead{(nJy)} &
\colhead{(nJy)} & \colhead{(nJy)} & \colhead{(nJy)} & \colhead{(nJy)} &
\colhead{(nJy)} & \colhead{(nJy)}
}
\startdata
398654 & \nodata &
$0.1\pm1.1$ & $1.6\pm1.1$ & $15.8\pm0.9$ & $16.5\pm1.8$ &
$17.4\pm1.4$ & $14.6\pm1.3$ & $12.2\pm1.3$ & $18.8\pm2.8$ &
$13.8\pm1.1$ & $15.9\pm1.7$ & $11.8\pm1.6$ & $16.9\pm1.0$ &
$15.1\pm2.0$ & $17.7\pm1.5$ \\
401961 & total &
$0.7\pm0.8$ & $-0.6\pm0.8$ & $21.4\pm1.3$ & \nodata &
\nodata & $22.6\pm1.7$ & \nodata & \nodata &
$22.3\pm1.5$ & $21.5\pm1.3$ & $17.2\pm1.3$ & $21.0\pm0.9$ &
$24.5\pm1.1$ & $22.0\pm0.8$ \\
401961 & comp\_1 &
$0.7\pm0.7$ & $-0.2\pm0.7$ & $17.7\pm0.7$ & \nodata &
\nodata & $22.0\pm0.6$ & \nodata & \nodata &
$18.6\pm0.5$ & $19.8\pm1.2$ & $15.9\pm1.1$ & $17.3\pm0.5$ &
$22.3\pm1.0$ & $19.6\pm0.7$ \\
401961 & comp\_2 &
$0.0\pm0.4$ & $-0.3\pm0.3$ & $3.7\pm1.0$ & \nodata &
\nodata & $0.6\pm1.6$ & \nodata & \nodata &
$3.7\pm1.5$ & $1.8\pm0.7$ & $1.2\pm0.7$ & $3.6\pm0.8$ &
$2.3\pm0.6$ & $2.4\pm0.5$ \\
424838 & total &
$-0.6\pm0.5$ & $0.3\pm0.5$ & $14.6\pm0.5$ & $17.7\pm1.2$ &
$19.9\pm1.0$ & $18.8\pm0.5$ & \nodata & \nodata &
$16.8\pm0.5$ & \nodata & $15.5\pm1.2$ & $14.4\pm0.5$ &
$15.8\pm1.1$ & $13.1\pm0.8$ \\
424838 & comp\_1 &
$-0.1\pm0.5$ & $0.3\pm0.4$ & $13.5\pm0.4$ & $17.1\pm1.1$ &
$18.0\pm0.9$ & $17.4\pm0.4$ & \nodata & \nodata &
$15.0\pm0.4$ & \nodata & $14.0\pm1.0$ & $13.2\pm0.4$ &
$14.3\pm0.9$ & $12.2\pm0.6$ \\
424838 & comp\_2 &
$-0.4\pm0.2$ & $0.0\pm0.2$ & $1.2\pm0.2$ & $0.6\pm0.5$ &
$1.9\pm0.4$ & $1.4\pm0.2$ & \nodata & \nodata &
$1.8\pm0.3$ & \nodata & $1.5\pm0.7$ & $1.2\pm0.3$ &
$1.5\pm0.6$ & $0.8\pm0.5$ \\
428365 & total &
$0.3\pm0.6$ & $-1.6\pm0.8$ & $15.9\pm0.9$ & \nodata &
\nodata & $15.8\pm0.9$ & \nodata & \nodata &
$13.1\pm1.1$ & \nodata & $14.1\pm2.2$ & $11.0\pm1.3$ &
$9.9\pm2.2$ & $11.8\pm1.5$ \\
428365 & comp\_1 &
$0.1\pm0.4$ & $-0.7\pm0.6$ & $7.9\pm0.6$ & \nodata &
\nodata & $10.0\pm0.7$ & \nodata & \nodata &
$7.7\pm0.8$ & \nodata & $7.1\pm1.6$ & $6.1\pm1.0$ &
$6.1\pm1.6$ & $9.1\pm1.1$ \\
428365 & comp\_2 &
$0.2\pm0.5$ & $-0.9\pm0.6$ & $8.0\pm0.7$ & \nodata &
\nodata & $5.8\pm0.6$ & \nodata & \nodata &
$5.4\pm0.8$ & \nodata & $7.1\pm1.5$ & $4.9\pm0.9$ &
$3.8\pm1.5$ & $2.7\pm1.0$ \\
441080 & total &
$-0.8\pm0.9$ & $0.7\pm0.9$ & $20.9\pm0.9$ & \nodata &
\nodata & $22.3\pm1.0$ & \nodata & \nodata &
$18.6\pm0.9$ & \nodata & $17.5\pm1.7$ & $17.6\pm1.0$ &
$16.4\pm1.6$ & $15.2\pm1.2$ \\
441080 & comp\_1 &
$0.6\pm0.6$ & $0.1\pm0.6$ & $13.8\pm0.6$ & \nodata &
\nodata & $15.6\pm0.7$ & \nodata & \nodata &
$12.8\pm0.6$ & \nodata & $9.9\pm1.1$ & $10.9\pm0.7$ &
$10.9\pm1.1$ & $11.9\pm0.9$ \\
441080 & comp\_2 &
$-1.4\pm0.7$ & $0.7\pm0.7$ & $7.1\pm0.7$ & \nodata &
\nodata & $6.7\pm0.8$ & \nodata & \nodata &
$5.8\pm0.7$ & \nodata & $7.6\pm1.3$ & $6.7\pm0.8$ &
$5.5\pm1.2$ & $3.3\pm0.9$ \\
452238 & total &
$-1.2\pm0.9$ & $1.3\pm0.7$ & $17.2\pm0.8$ & $22.3\pm1.4$ &
$23.5\pm1.1$ & $24.0\pm0.8$ & \nodata & \nodata &
$22.7\pm0.8$ & $19.4\pm1.5$ & $22.7\pm1.3$ & $20.7\pm0.7$ &
$26.9\pm1.5$ & $25.7\pm1.0$ \\
452238 & comp\_1 &
$-1.0\pm0.7$ & $0.8\pm0.6$ & $11.9\pm0.6$ & $17.5\pm1.2$ &
$16.6\pm0.9$ & $18.0\pm0.7$ & \nodata & \nodata &
$17.9\pm0.7$ & $14.6\pm1.3$ & $18.7\pm1.0$ & $16.3\pm0.6$ &
$22.3\pm1.1$ & $20.9\pm0.8$ \\
452238 & comp\_2 &
$-0.2\pm0.5$ & $0.5\pm0.4$ & $5.3\pm0.5$ & $4.8\pm0.8$ &
$6.9\pm0.6$ & $6.0\pm0.5$ & \nodata & \nodata &
$4.8\pm0.5$ & $4.7\pm0.8$ & $4.0\pm0.8$ & $4.5\pm0.4$ &
$4.6\pm0.9$ & $4.8\pm0.6$ \\
462180 & total &
$-2.0\pm1.1$ & $3.1\pm1.1$ & $34.7\pm1.0$ & $41.6\pm1.9$ &
$36.4\pm1.5$ & $33.5\pm1.3$ & $28.8\pm1.4$ & $31.7\pm2.9$ &
$31.2\pm1.3$ & $26.4\pm2.1$ & $26.7\pm1.8$ & $28.7\pm1.2$ &
$27.9\pm2.4$ & $32.4\pm1.8$ \\
462180 & comp\_1 &
$-0.9\pm0.8$ & $1.8\pm0.7$ & $19.0\pm0.6$ & $23.2\pm1.3$ &
$19.7\pm1.0$ & $18.3\pm0.9$ & $16.4\pm0.9$ & $16.9\pm1.7$ &
$15.3\pm0.8$ & $13.6\pm1.3$ & $13.6\pm1.0$ & $15.6\pm0.7$ &
$16.4\pm1.4$ & $17.3\pm1.0$ \\
462180 & comp\_2 &
$0.0\pm0.5$ & $0.4\pm0.5$ & $9.5\pm0.4$ & $9.4\pm0.9$ &
$9.6\pm0.7$ & $8.1\pm0.6$ & $6.6\pm0.7$ & $7.5\pm1.4$ &
$7.9\pm0.6$ & $6.3\pm1.1$ & $6.3\pm0.9$ & $6.3\pm0.6$ &
$8.1\pm1.2$ & $7.7\pm0.9$ \\
462180 & comp\_3 &
$0.3\pm0.4$ & $0.1\pm0.4$ & $3.7\pm0.3$ & $4.0\pm0.7$ &
$3.6\pm0.6$ & $3.6\pm0.5$ & $2.9\pm0.6$ & $3.1\pm1.2$ &
$3.8\pm0.5$ & $3.5\pm0.8$ & $3.9\pm0.8$ & $3.8\pm0.5$ &
$2.3\pm1.1$ & $4.5\pm0.8$ \\
462180 & comp\_4 &
$-1.4\pm0.5$ & $0.8\pm0.5$ & $2.5\pm0.4$ & $5.0\pm0.9$ &
$3.4\pm0.7$ & $3.4\pm0.6$ & $3.0\pm0.6$ & $4.2\pm1.4$ &
$4.2\pm0.6$ & $3.0\pm0.9$ & $2.8\pm0.9$ & $3.0\pm0.6$ &
$1.1\pm1.2$ & $2.9\pm0.9$ \\
462894 & total &
$1.7\pm0.9$ & $-0.5\pm0.7$ & $35.4\pm0.8$ & $41.0\pm1.5$ &
$41.3\pm1.2$ & $44.0\pm0.9$ & \nodata & \nodata &
$36.7\pm1.0$ & $31.2\pm3.5$ & $33.4\pm1.5$ & $34.6\pm0.8$ &
$38.8\pm1.7$ & $37.2\pm1.1$ \\
462894 & comp\_1 &
$1.6\pm0.7$ & $-0.7\pm0.5$ & $25.1\pm0.7$ & $28.5\pm1.2$ &
$29.8\pm1.0$ & $30.7\pm0.7$ & \nodata & \nodata &
$26.1\pm0.8$ & $23.0\pm3.2$ & $23.2\pm1.2$ & $25.3\pm0.7$ &
$29.7\pm1.3$ & $25.6\pm0.9$ \\
462894 & comp\_2 &
$0.1\pm0.6$ & $0.2\pm0.4$ & $10.3\pm0.5$ & $12.5\pm0.9$ &
$11.6\pm0.7$ & $13.4\pm0.5$ & \nodata & \nodata &
$10.6\pm0.6$ & $8.2\pm1.2$ & $10.2\pm0.9$ & $9.4\pm0.5$ &
$9.1\pm1.0$ & $11.6\pm0.7$ \\
463424 & \nodata &
$0.2\pm0.7$ & $1.4\pm0.7$ & $9.4\pm0.7$ & \nodata &
\nodata & $10.1\pm0.7$ & \nodata & \nodata &
$9.0\pm0.6$ & \nodata & $8.4\pm1.2$ & $9.5\pm0.8$ &
$10.0\pm1.2$ & $16.0\pm0.9$ \\
463518 & \nodata &
$0.0\pm0.3$ & $0.0\pm0.3$ & $16.9\pm0.4$ & $23.2\pm0.7$ &
$15.7\pm0.6$ & $16.2\pm0.4$ & \nodata & \nodata &
$16.9\pm0.5$ & $17.6\pm2.6$ & $23.4\pm0.8$ & $22.7\pm0.4$ &
$26.0\pm0.8$ & $23.5\pm0.6$ \\
469153 & \nodata &
$0.1\pm0.3$ & $0.1\pm0.2$ & $17.3\pm0.3$ & $21.0\pm0.6$ &
$21.9\pm0.6$ & $21.5\pm0.3$ & \nodata & \nodata &
$22.5\pm0.4$ & $18.2\pm3.2$ & $21.6\pm0.8$ & $23.2\pm0.4$ &
$27.1\pm0.7$ & $37.0\pm0.5$ \\
469779 & \nodata &
$0.7\pm0.6$ & $-0.1\pm0.6$ & $13.8\pm0.6$ & $28.6\pm1.2$ &
$29.6\pm0.9$ & $28.3\pm0.4$ & $27.5\pm0.9$ & $28.4\pm1.7$ &
$24.4\pm0.8$ & $19.1\pm2.5$ & $22.4\pm1.1$ & $20.8\pm0.7$ &
$21.3\pm1.4$ & $22.9\pm1.1$ \\
470387 & \nodata &
$-0.2\pm0.4$ & $0.2\pm0.3$ & $16.3\pm0.4$ & $18.9\pm0.7$ &
$20.8\pm0.6$ & $19.4\pm0.4$ & \nodata & \nodata &
$17.4\pm0.4$ & $13.6\pm2.3$ & $16.5\pm0.8$ & $16.1\pm0.5$ &
$18.9\pm0.8$ & $18.6\pm0.6$ \\
470678 & \nodata &
$0.2\pm0.5$ & $2.3\pm0.5$ & $7.3\pm0.5$ & \nodata &
\nodata & $8.3\pm0.6$ & \nodata & $10.3\pm1.3$ &
$11.1\pm0.8$ & $11.2\pm1.5$ & $15.2\pm1.0$ & $16.3\pm0.6$ &
$21.1\pm1.4$ & $44.9\pm1.1$ \\
495517 & \nodata &
$-0.7\pm0.6$ & $-0.2\pm0.6$ & $7.5\pm0.5$ & $12.5\pm0.8$ &
$11.0\pm0.7$ & $12.0\pm0.7$ & $10.5\pm0.6$ & $11.1\pm1.4$ &
$9.9\pm0.6$ & $9.5\pm0.9$ & $9.2\pm0.9$ & $9.6\pm0.5$ &
$9.2\pm1.1$ & $8.6\pm0.9$ \\
498913 & \nodata &
$-1.2\pm0.4$ & $0.3\pm0.4$ & $8.5\pm0.5$ & $10.3\pm1.0$ &
$12.3\pm0.8$ & $10.3\pm1.1$ & $11.3\pm0.8$ & $12.6\pm1.2$ &
$11.8\pm0.5$ & $11.4\pm1.2$ & $12.4\pm1.0$ & $13.0\pm0.7$ &
$13.7\pm1.1$ & $14.1\pm1.3$ \\
499935 & \nodata &
\nodata & \nodata & $11.5\pm1.0$ & $12.0\pm1.0$ &
$13.4\pm0.8$ & \nodata & $12.1\pm0.8$ & $15.3\pm1.5$ &
$13.5\pm0.9$ & $13.2\pm1.3$ & $13.0\pm1.0$ & $12.6\pm0.7$ &
$12.2\pm1.7$ & $13.4\pm1.3$ \\
500345 & \nodata &
$-0.1\pm0.3$ & $1.0\pm0.5$ & $16.1\pm0.6$ & \nodata &
\nodata & $16.5\pm0.6$ & \nodata & \nodata &
$15.9\pm0.6$ & \nodata & $12.7\pm1.1$ & $14.8\pm0.7$ &
$19.9\pm1.1$ & $25.3\pm0.8$ \\
503989 & \nodata &
$1.2\pm0.7$ & $1.2\pm0.7$ & $11.2\pm0.6$ & $10.9\pm0.8$ &
$10.6\pm0.7$ & $8.7\pm0.8$ & $10.6\pm0.6$ & $11.5\pm1.3$ &
$9.7\pm0.8$ & $9.5\pm1.2$ & $9.6\pm0.9$ & $11.3\pm0.7$ &
$9.9\pm1.5$ & $13.3\pm1.2$ \\
512976 & \nodata &
$0.1\pm0.4$ & $0.5\pm0.4$ & $8.5\pm0.4$ & $11.0\pm0.7$ &
$10.4\pm0.6$ & $9.4\pm0.6$ & $9.4\pm0.6$ & $9.8\pm1.2$ &
$7.6\pm0.7$ & $8.1\pm0.8$ & $7.1\pm0.9$ & $8.1\pm0.6$ &
$5.7\pm1.4$ & $5.7\pm1.1$ \\
\enddata
\tablecomments{
Fluxes are derived from \texttt{ForcePho} fitting with \sersic\ profiles across all available NIRCam bands. For galaxies with multiple structural components, fluxes for each component are listed separately.
}
\end{deluxetable*}

\renewcommand{\thefigure}{A}
\begin{figure*}
    \centering
    \includegraphics[width=0.65\linewidth]{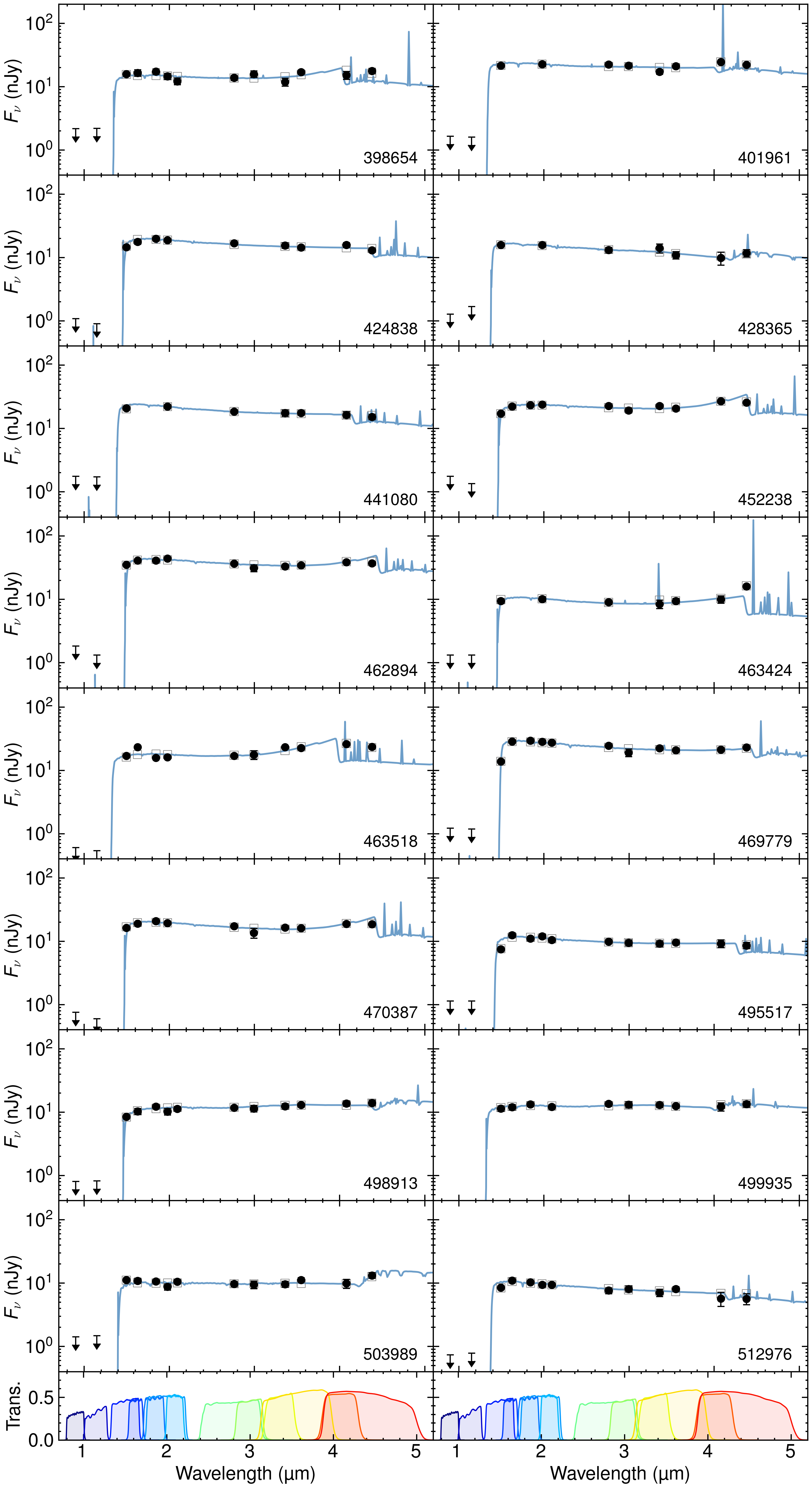}
    \caption{SEDs of the other objects in the overdensity, in the same format as Figure~\ref{fig:balmer_break}. Black points show the JWST/NIRCam photometry, with non-detections shown as $2\sigma$ upper limits, and blue curves show the best-fit {\tt Prospector} models. The lower panels show the NIRCam filter transmission curves. }
    \label{fig:sed_all}
\end{figure*}

\section{Systemic redshift estimate}

To estimate the systemic redshift of the overdensity, we combine the redshift posterior probability distribution functions (PDFs) of the 18 member candidates under the assumption that most of them have a  common redshift, while allowing a small number of projected interlopers. 

Let $D_i$ denote the photometric data of galaxy $i$, and let $p_i(z)\equiv p(z\mid D_i)$ be its redshift posterior PDF from the {\tt Prospector} SED fitting. Because the individual redshift priors are uniform, $p_i(z)$ is proportional to the likelihood as a function of $z$. We assume that true members share a common redshift $z_{\rm sys}$, with negligible intrinsic dispersion compared to the width of the photometric-redshift PDFs. We further allow exactly $K$ outliers among the $N=18$ galaxies. For a given $K$, the marginalized likelihood is
\begin{equation}
p(\{D_i\}\mid z_{\rm sys},K)\propto
\frac{1}{\binom{N}{K}}
\sum_{S:\,|S|=N-K}\prod_{i\in S} p_i(z_{\rm sys}),
\end{equation}
where the sum runs over all subsets $S$ of size $N-K$ interpreted as the member set. The factor $1/\binom{N}{K}$ assigns equal prior weight to each choice of the $K$ outliers. In this expression, the outlier contribution has been marginalized over redshift and absorbed into a normalization constant, since it does not depend on $z_{\rm sys}$ for a uniform redshift prior.

We place a Poisson prior on the outlier number,
\begin{equation}
p(K)\propto \frac{\lambda^K e^{-\lambda}}{K!},\qquad K=0,\dots,N,
\end{equation}
with $\lambda=4$, given the expected number of projected interlopers (Section~\ref{sec:overdensity}). The Poisson prior is truncated to the allowed range $0\le K\le N$ and renormalized. The posterior for the systemic redshift is then
\begin{equation}
p(z_{\rm sys}\mid \{D_i\}) \propto
\sum_{K=0}^{N} p(\{D_i\}\mid z_{\rm sys},K)\,p(K).
\end{equation}

The joint posterior yields a median systemic redshift of $z_{\rm sys}= 10.41$, with the 16th and 84th percentiles at 10.03 and 10.87, respectively, as shown in Figure~\ref{fig:photoz}.

\renewcommand{\thefigure}{B}
\begin{figure}
    \centering
    \includegraphics[width=0.9\linewidth]{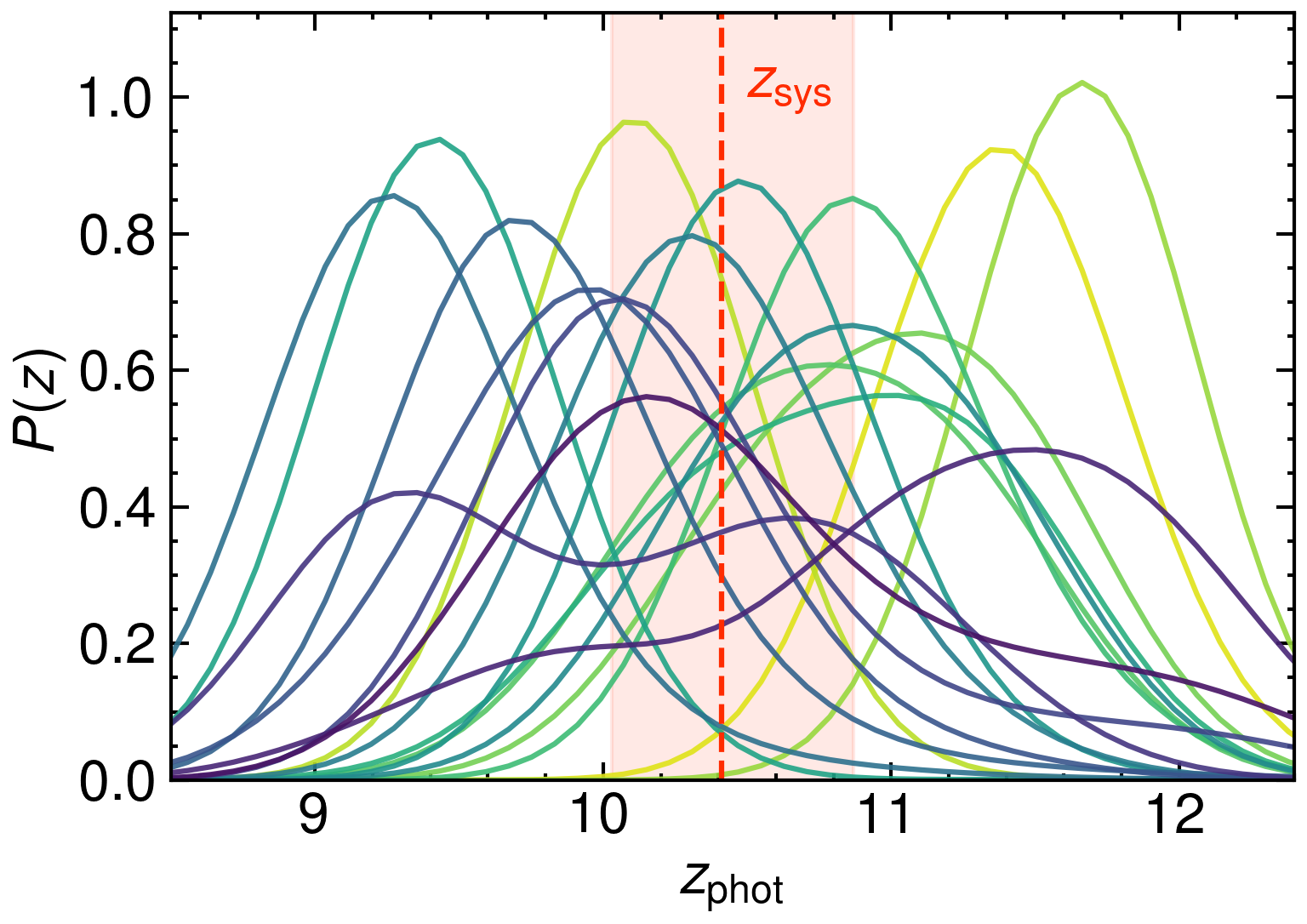}
    \caption{Individual redshift posterior PDFs for galaxies in the overdensity. The dashed vertical line marks the estimated common-redshift  $z_\mathrm{sys}=10.4$, inferred from their joint posterior while allowing for a small number of interlopers. The shaded band shows the 16th--84th percentile range.}
    \label{fig:photoz}
\end{figure}

\section{Candidates Excluded from the Fiducial Sample}
\label{sec:dubious}
{\bf ID 423134} (RA $=$ 53.011013, Dec $=-27.833990$) This source has a nominal $1.49\pm0.45$\,nJy ($3.3\sigma$) F115W flux in the JADES DR5 catalog in aperture photometry. However, the measurement is contaminated by a nearby luminous galaxy at $z_\mathrm{phot}=2.88$. The F115W image shows no compact emission associated with the source. Using aperture photometry with a 0.1$''$ radius and subtract a local background estimated from an annulus with inner and outer radii of 0.15$''$ and 0.2$''$, we measure an F115W flux of $0.4\pm0.8$\,nJy. The uncertainty includes the formal pixel-level measurement error within the source aperture, as well as contributions from background fluctuations and the uncertainty in the estimated background level, both quantified using the pixel standard deviation within the background annulus. This non-detection thus may imply a Ly$\alpha$ dropout signature at $z>10$, making the source a plausible member of the overdensity. Nonetheless, we note that the nearby galaxy happens to have a Balmer break between F115W and F150W, raising concern that the candidate may be associated with this lower-redshift galaxy. Given this possibility, we exclude this source from our fiducial sample. However, this scenario would require an unusually strong Balmer break to explain the non-detection. Comparing with the nearby galaxy with a flux ratio of F115W/F200W $=$ 0.56, this source has a Bayesian $84\%$ upper limit of F115W($<84\%$)/F200W $=$ 0.11, assuming a non-negative uniform prior following Eisenstein et al. in prep. Such strong Balmer breaks are rare, especially in faint galaxies.

{\bf ID 462180} (RA $=$ 52.963890, Dec $=$ $-27.817278$)
This source is marginally detected in F115W at $2.9\sigma$ in {\tt ForcePho} photometry. The emission is on the outskirts of the galaxy, making the reality of the flux uncertain. The object contains four subcomponents with consistent photometric redshifts. The SED modeling with {\tt Prospector} favors a solution at $z\approx9$ due to the tentative F115W flux. This source has a total stellar mass of $\log\,(M/M_\odot)=8.7^{+1.4}_{-2.6}$ combining all four components.

{\bf ID 470678} (RA $=$ 52.992997, Dec $=$ $-27.842358$)
This source is marginally detected in F115W at $2.7\sigma$ in {\tt ForcePho} photometry. The tentative emission is compact and centered on the source, suggesting that it may be real. The SED is unusually red, as shown in Figure~\ref{fig:sed_dubious}. The inferred UV slope is $\beta=-0.7\pm0.2$, significantly redder than that of typical high-redshift galaxies. The SED fitting favors substantial dust attenuation with $A_V=1.06$ and a large stellar mass of $\log\,(M/M_\odot)=8.9$. Alternatively, it could be an AGN. This source is detected in MIRI/F770W at $114\pm32$\,nJy, consistent with a power-law extrapolation from the rest-frame UV flux.

\renewcommand{\thefigure}{C}
\begin{figure*}
    \centering
    \includegraphics[width=0.6\linewidth]{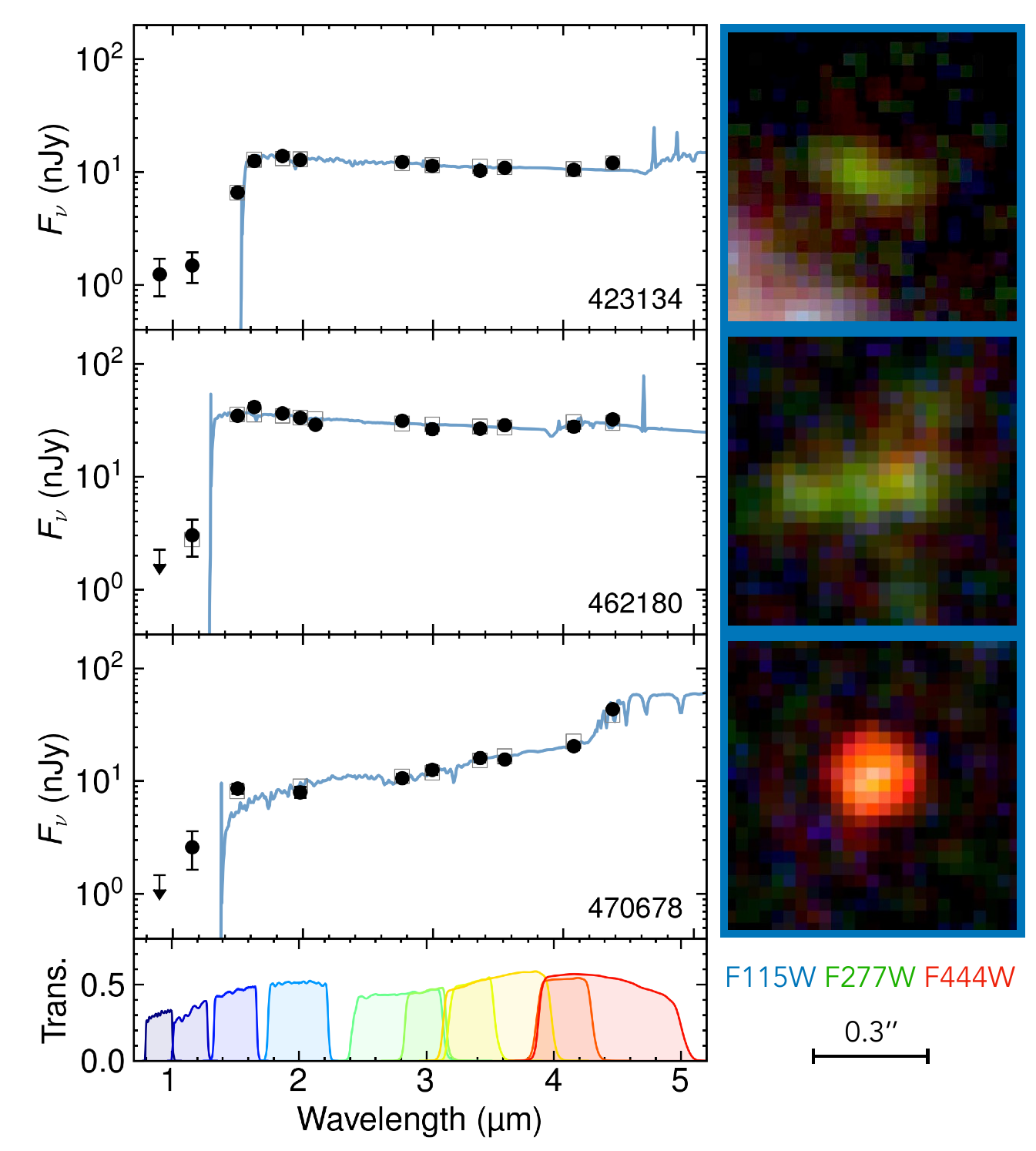}
    \caption{SEDs and NIRCam cutout images of the three candidates excluded from the fiducial overdensity sample. Black points show the photometry, with non-detections plotted as $2\sigma$ upper limits, and the blue curves show the best-fit {\tt Prospector} models. The lower panel shows the NIRCam filter transmission curves. The cutouts on the right use the F115W, F277W, and F444W bands as blue, green, and red, respectively. JADES-GS-470678 is additionally detected in MIRI/F770W at $114\pm32$\,nJy.}
    \label{fig:sed_dubious}   
\end{figure*}

\bibliography{reference}{}
\bibliographystyle{aasjournalv7}

\end{document}